\documentclass[prb, showpacs, superscriptaddress, amsmath,amssymb,aps, 
twocolumn,10pt]{revtex4-1}
\usepackage{txfonts}
\usepackage{empheq}
\usepackage{bbm}

\usepackage{amsmath}%
\usepackage{amssymb}%
\usepackage{graphicx}% Include figure files
\usepackage{dcolumn}% Align table columns on decimal point
\usepackage{bm}% bold math
\usepackage{algorithmic}%
\usepackage{color}%
\usepackage[normalem]{ulem}
\usepackage{ifpdf}

\begin{document}

\title{Cubic-scaling iterative solution of the Bethe-Salpeter equation for finite systems}
\author{M. P. Ljungberg}
\affiliation{Deparment of Physics, Phillips-University Marburg, 
Renthof 5, 35032 Marburg, Germany}
\affiliation{Donostia International Physics Center, 
Paseo Manuel de Lardizabal, 4. E-20018 Donostia-San Sebasti\'{a}n, Spain}

\author{P. Koval}
\affiliation{Donostia International Physics Center, 
Paseo Manuel de Lardizabal, 4. E-20018 Donostia-San Sebasti\'{a}n, Spain}

\author{F. Ferrari}
\affiliation{Donostia International Physics Center, 
Paseo Manuel de Lardizabal, 4. E-20018 Donostia-San Sebasti\'{a}n, Spain}
 \affiliation{Dipartimento di Scienza dei Materiali, Universit\`{a} degli Studi di Milano-Bicocca, I-20125, Milano, Italy.}

\author{D. Foerster}
\affiliation{LOMA, Universit\'e de Bordeaux 1, 
351 Cours de la Liberation, 33405 Talence, France}

\author{D. S\'anchez-Portal}
\affiliation{Centro de F\'{\i}sica de Materiales CFM-MPC,
Centro Mixto CSIC-UPV/EHU, Paseo Manuel de Lardizabal 5, E-20018 
San Sebasti\'an, Spain}
\affiliation{Donostia International Physics Center, 
Paseo Manuel de Lardizabal, 4. E-20018 Donostia-San Sebasti\'{a}n, Spain}

\begin{abstract}
The Bethe-Salpeter equation (BSE) is currently the state of the art in 
the description of neutral electron excitations in both solids and large finite systems.
It is capable of accurately treating charge-transfer excitations that present difficulties for simpler approaches. 
We present a local basis set formulation of the BSE for molecules where the optical spectrum is computed with the iterative Haydock recursion scheme, leading to a low computational
complexity and memory footprint. Using a variant of the algorithm we can go beyond the Tamm-Dancoff approximation (TDA). We rederive the recursion relations for general matrix elements of a resolvent, show how they translate into continued fractions, and study the convergence of the method with the number of recursion coefficients and the role of different terminators.
Due to the locality of the basis functions the computational cost of each iteration 
scales asymptotically as $O(N^3)$ with the number of atoms, while 
the number of iterations is typically much lower than the size of the 
underlying electron-hole basis. In practice we see that , even for systems with thousands of orbitals, 
the runtime will be dominated by the $O(N^2)$ operation of applying the Coulomb 
kernel in the atomic orbital representation
\end{abstract}

\maketitle

\section{Introduction}

%%%
%%% Motivation
%%%

{\it Ab initio} simulation of optical spectra is an essential tool 
in the study of excited state electronic properties of solids, 
molecules and nanostructures. For finite systems time-dependent 
density functional theory (TDDFT) \cite{Runge_Gross:1984}
based on local or semi local functionals is  widely used.
However, TDDFT fails in certain cases, notably for charge
transfer excitations \cite{Casida_inbook:2006} which are essential
in, e.g., photovoltaic applications. An alternative to TDDFT
is Hedin's $GW$ approximation \cite{Hedin:1965} followed by the solution
of the Bethe-Salpeter equation (BSE) \cite{Onida_Reining:2002}.
Based on many-body perturbation theory \cite{Fetter_Walecka,Abrikosov:1963}, 
the $GW$/BSE method is a more systematic approach than TDDFT, and it has 
been shown to give a qualitatively correct description of 
excitonic effects in solids \cite{Hanke-Sham:1980,Onida_Reining:2002}
and charge transfer excitations\cite{Bechstedt:1997,Blase:2011}.

The Bethe-Salpeter equation is a Dyson-like equation for the two-particle Green's function, or equivalently for the four-point polarizability \cite{Reining:1997}. 
Within the field of electronic structure theory,  developments of the BSE can be traced back to the beginning of sixties \cite{Abrikosov:1963, Nozieres:1964, Sham-Rice:1966}, with 
 the first \emph{ab initio} implementations appearing a couple of decades later \cite{Louie:1993,Reining:1995,Benedict-Shirley:1998}. The GW/BSE method has been implemented using plane waves and real space grids,  \cite{Reining:1997,Benedict:2003,Bechstedt:2004, 
Thygesen:2011,Rocca:2010,Gruening-Gonze:2011,
Kresse-Bechstedt:2008,Gatti-Sottile:2013,Shirley:1999,Kronik-Chelikowsky:2007}, linear combination of atomic orbitals (LCAO) \cite{Rohlfing-Louie:1998, Casida:2005,Rohlfing:2012,
Bruneval:2013, Blase:2014}
and within the FLAPW framework \cite{AmbroschDraxl:2006}.  
In practice, the standard way of solving the BSE is by converting it to an effective eigenvalue problem in a particle-hole basis. 
Since the size of the particle-hole basis scales quadratically with the number of atoms $N$, a straightforward diagonalization of the BSE Hamiltonian will scale like $O(N^6)$. This very steep scaling makes it difficult to treat large scale systems like nanostructures and realistic models of organic photovoltaic devices. For such systems an improved scaling with the number of atoms would be highly beneficial.

Avoiding an explicit diagonalization of the BSE Hamiltonian can be done by using an iterative method to obtain a few low-lying transitions (e.g. the Davidsson method \cite{Bai:2000, Valiev:2010}), or to directly aim for the spectrum, which can be done frequency by frequency using for example the GMRES method \cite{saad, Bai:2000,Koval:2010} or for the full spectrum with the Haydock recursion scheme \cite{Haydock:1972, Rocca:2008, Gruening-Gonze:2011}. Another option is to go over to the time domain and solve the equations of motion by time propagation \cite{Bechstedt:2003, Marini:TP:2011}. These methods only require matrix-vector products to be performed, and assuming that the number of iterations, or time steps, is much smaller than the size of the particle-hole basis, the asymptotic scaling will be $O(N^4)$. However, setting up the BSE Hamiltonian explicitly will still have the cost of  $O(N^5)$, and to avoid this, the matrix-vector products need to be performed on the fly, without explicitly constructing the matrix. 

Benedict and Shirley made use of the Haydock recursion method to compute optical spectra in the Tamm-Dancoff approximation (TDA) without actually computing the whole BSE Hamiltonian \cite{Shirley:1999}. 
This was achieved by using, in addition to the particle-hole basis, a real space grid product basis $| \bm{x}, \bm{y} \rangle$, in which the screened direct Coulomb interaction is diagonal (the exchange term is sparse in this representation). The scaling of the algorithm was reported to be $O(N^4)$ with the number of atoms, however, a more careful analysis shows that it can be made to scale like  $O(N^3)$ by a proper ordering of the loops \cite{Benedict:unpublished}. 

This favorable scaling is heavily based on the use of a real-space representation for the particle-hole states. Similar gains can be obtained with the use of LCAO basis sets, where the same asymptotic scaling can be obtained by making use of the sparsity in both direct and exchange Coulomb interaction terms.
It should be mentioned that by using additional assumptions of locality, which implies screening away Coulomb matrix elements between basis functions that are spatially far from each other, one could even achieve linear scaling\cite{Ochsenfeld:2007}, however, the BSE has so far not been treated with these methods. Another linear scaling approach to many-body theory methods has recently been published by Baer and coworkers that make use of stochastic wave functions together with time propagation \cite{Baer:RPA:2013,Baer:GW:PRL:2014,Baer:BSE:2015}.

In the present publication, we will not venture into the realm of linear scaling but rather make use of the more standard iterative methods that, together with locality, lead to cubic scaling with the number of atoms.  We present an iterative algorithm to obtain the BSE spectrum for molecules, making use of localized basis sets both for orbitals and products of orbitals. To go beyond the TDA a pseudo-Hermitian version the Haydock recursion scheme \cite{Gruening-Gonze:2011} is used. We derive the recursion relations for general matrix elements of a resolvent and show how they translate into continued fractions. Our method has been interfaced to the SESTA code \cite{SIESTA} which is widely used for ground state density functional theory calculations (as an alternative, we can do all-electron calculation using numerical orbitals in an in-house implementation). For the case of the benzene molecule, as a prototypical example, we present a detailed study of the convergence properties of the iterative method, both within the Tamm-Dancoff approximation and for the full BSE. In particular, we study the effect of different termination schemes. Furthermore, for the sake of clarity, we provide a detailed account of the BSE method itself using our notation. Our algorithm scales asymptotically like $O(N^3)$ with the number of atoms and uses $O(N^2)$ memory. We present proof of principle calculations of our implementation, where the runtime is seen to be dominated by the $O(N^2)$ scaling operations for systems up to several thousand orbitals,  and discuss some of the bottlenecks and possible improvements of the scheme. 
 
%%%
%%% Algebra of the implementation
%%%
\section{Theory}

\subsection{Quasiparticles with the $GW$ approximation}
Before the BSE can be set up and solved, the quasiparticle energies must be obtained from 
a preceding $GW$ calculation \cite{Hedin:1965}. Since the details of our $GW$ implementation
have been published elsewhere \cite{Foerster:2011, Koval:2014}, we will
here only give a brief summary of the method.
The poles of the one-particle Green's function $G$ for an $N$-electron 
system occur at the ground and excited states 
of the corresponding  $N$+1 and $N$-1 systems, that is at the electron 
addition and removal energies.  Hedin's $GW$ approximation
connects the (irreducible) polarizability $P$, the 
non-interacting and interacting Green's functions ($G^0$ and $G$), 
the screened interaction $W$, and the self energy $\Sigma$
in a set of closed equations
\begin{align}
 P(\bm{r},\bm{r}',\omega) &= i \int G^0(\bm{r},\bm{r}',\omega-\omega')
 G^0(\bm{r}',\bm{r},\omega') d\omega' \label{eq:pol-space},\\ 
\nonumber W(\bm{r},\bm{r}', \omega ) &=  v(\bm{r},\bm{r}') + \\
 &\int v(\bm{r},\bm{r}_2) P(\bm{r}_2,\bm{r}_3, \omega) 
 W(\bm{r}_3,\bm{r}', \omega) d^3r_2d^3r_3,\label{scr-inter-rspace}\\
 \Sigma(\bm{r},\bm{r}',\omega) &= 
 \frac{i}{2\pi} \int G^0(\bm{r},\bm{r}', \omega' ) 
 W(\bm{r},\bm{r}', \omega -\omega') d\omega' , \label{se-rspace}\\
\nonumber G(\bm{r},\bm{r}', \omega ) &= G^0(\bm{r},\bm{r}', \omega ) + \\
&\int G^0(\bm{r},\bm{r}_2, \omega ) \Sigma(\bm{r}_2,\bm{r}_3,\omega) G(\bm{r}_3,\bm{r}', \omega )
d^3r_2d^3r_3. \label{eq:Dyson_eq}
%\label{eq:GW_eqs}
\end{align}

In our implementation
of the $GW$ method the Green's function is expanded in a basis of 
numerical atomic orbitals (AO) of finite support $\{ f_{a}(\bm{r})\}$
\begin{equation}
G(\bm{r},\bm{r}',\omega) = \sum_{aa'bb'}
f_{a}(\bm{r}) S^{-1}_{aa'}G_{a'b'}(\omega) S^{-1}_{b'b}f^*_{b}(\bm{r}').
\label{gf-lcao}
\end{equation}
Here and in the following we explicitly write out the overlaps
$S_{ab} = \int f^*_a(\bm{r}) f_b(\bm{r}) d^3r$ when they appear, the matrix
quantities $G_{ab}(\omega)$ are always contravariant
and the placement of the indices as subscripts or superscript is arbitrary. 
With this representation of the Green's function $G$, we see that
the polarizability (\ref{eq:pol-space}) involves products of 
AOs $f_{a}(\bm{r})f^*_{b}(\bm{r})$. These products are expanded in an 
(auxiliary) product basis $\{ F_{\mu}(\bm{r})\}$ of localized
numerical functions \cite{ Foerster:2011, Koval:2014} 
\begin{equation}
\begin{split}
f_a(\bm{r}) f^*_b(\bm{r}) &=  \sum_{\mu} V^{ab}_{\mu} F_{\mu}(\bm{r}) \,,
\end{split}
\label{eq:ao_vertex}
\end{equation}
where the expansion coefficients $V^{ab}_{\mu}$ and the product 
basis functions $\{F_{\mu}(\bm{r})\}$ are determined by numerically
expanding the products around a common center and removing redundant 
functions by a diagonalization based procedure \cite{df:2008}. Only overlapping pairs of orbitals
are considered, making the matrix of expansion coefficients sparse when using AOs of local support. 
The indices $\{ aa'bb'\}$ will be reserved for 
 atomic orbitals and $\{ \mu, \nu\}$ for 
 product functions of atomic orbitals in the following. 
Using the product basis, the polarizability $P(\bm{r},\bm{r}',\omega)$
is represented similarly to the Green's functions (\ref{gf-lcao})
\begin{equation}
P(\bm{r},\bm{r}',\omega) = 
\sum_{\mu\mu'\nu\nu'}F_{\mu}(\bm{r}) 
S^{-1}_{\mu\mu'}P_{\mu'\nu'}(\omega) S^{-1}_{\nu'\nu}F^*_{\nu }(\bm{r}'),
\label{pol-ansatz}
\end{equation}
where the overlap of the product functions 
$S_{\mu \nu} =\int F^*_{\mu}(\bm{r}) F_{\nu}(\bm{r}) d^3r$ appears.
Similarly, it can be seen from equation (\ref{scr-inter-rspace}) that the matrix 
elements of the bare $v$ and screened $W$ Coulomb interaction must be expanded
in the product basis, while the self-energy $\Sigma$ is expanded in the AO basis. For 
finite systems both the $\{f_a(\bm{r})\}$ and the product basis $\{F_{\mu}(\bm{r})\}$ can 
be chosen as real. 

The frequency-dependent quantities like $G_{ab}(\omega)$
and $P_{\mu\nu}(\omega)$ are represented on an even-spaced, real-axis,
frequency grid via their corresponding spectral functions.
An imaginary part of the energy is added in the Green's function $G^0(\omega)$ 
and polarizability $P(\omega)$, that is sufficient to ensure their smoothness 
on the chosen frequency grid. The convolutions of spectral functions implied
by equations (\ref{eq:pol-space}) and (\ref{se-rspace}) are computed 
via fast Fourier transforms. Due to the the fast convolutions and the locality of the
product basis set, the asymptotic scaling of the algorithm 
is $O(N^3)$ with the number of atoms $N$ \cite{Foerster:2011}.
Finally the Dyson equation (\ref{eq:Dyson_eq})
is directly solved for each frequency to obtain $G_{ab}(\omega)$. The quasiparticle energies are poles in
$G_{ab}(\omega)$ and can in certain cases be determined 
from inspection of the density of states. This does not give the quasiparticle wave function, however.
In this paper we adopt the standard way of proceeding and assume that
the Kohn-Sham \cite{Kohn-Sham:1965} (KS)  or Hartree-Fock (HF) eigenfunctions that are used
to construct the zeroth order Green's function $G^0(\omega)$
are good approximations to
the quasiparticle states, so that they can be kept fixed and only the
quasiparticle energy corrected. We will here only consider the so-called
$G_0W_0$ approximation where a single iteration of the $GW$ equations is 
performed without  self-consistency. We focus on the KS ``starting point'' in this 
subsection. The KS Hamiltonian is
\begin{equation}
\begin{split}
H^{\text{KS}} = T + V^{\text{ext}}(\bm r) + V^\text{H} (\bm r)+  V^\text{xc} (\bm r)
\end{split}
%\label{eq:time_evolution}s
\end{equation}
with $T$ the kinetic energy,
$V^\text{ext}(\bm r)$ the external potential,
$V^\text{H} (\bm r)$ the Hartree potential and $V^\text{xc} (\bm r)$
the exchange-correlation potential. 
The KS eigenfunctions are expanded in the AO basis
\begin{equation}
\begin{split}
\psi_i(\bm{r}) = \sum_a X_{ia} f_a(\bm{r}) \, ,
\end{split}
\label{eq:MO_expansion}
\end{equation}
where $X_{ia} = \sum_{a'} S^{-1}_{aa'} \langle a' | i \rangle $ are 
the eigenvectors of the generalized eigenvalue problem 
\begin{equation}
\begin{split}
  \sum_{b} H^\text{KS}_{ab} X_{ib}  = \epsilon_i^{KS}  \sum_{b} S_{ab} X_{ib} \, .
\end{split}
%\label{eq:time_evolution}s
\end{equation}
If we additionally assume that interacting Green's function $G$ is diagonal 
in the KS eigenstates $\psi_i(\bm{r})$,  the Dyson equation (\ref{eq:Dyson_eq}) 
reduces to a set of scalar equations
\begin{equation}
G_{ii}(\omega) = \frac{1}{\omega - 
\epsilon^{\text{KS}}_i - (\Sigma_{ii}(\omega) -V^{\text{xc}}_{ii} )} \, ,
\label{eq:mo_vertex}
\end{equation}
where we have subtracted the exchange-correlation potential $V^{\text{xc}}_{ii}$
in order to be 
able to work with the KS eigenvalues. The (assumed real) poles are then found by identifying
the zeros of the denominator, either by a graphical solution if the full
frequency-dependent
quantities are available, or more commonly, by an expansion of		
 $\Sigma_{ii}(\omega)$ around $\epsilon^{\text{KS}}_i$, which leads to
 \begin{equation}
\begin{split}
\epsilon^{GW}_i &= \epsilon^{\text{KS}}_i + Z_i(\operatorname{Re} \Sigma_{ii}(\epsilon^{\text{KS}}_i ) -
V^{\text{xc}}_{ii} ) \, , \\
Z_i &= 
\left (1-\frac{\partial \operatorname{Re} \Sigma_{ii}(\omega)}{ \partial \omega} \Big |_{
\omega=\epsilon^{\text{KS}}_i} 
\right)^{-1} \, . 
\end{split}
\label{eq:quasiparticle_GW_Z_fac}
\end{equation}

Since we have access to the full frequency dependence of 
the self energy 
we can use the graphical method, which in principle is more accurate and 
also has the advantage 
that problems with satellite peaks can be avoided \cite{Ljungberg_BSE_SS:2015}. For comparison 
purposes we will also make use of the simpler equation (\ref{eq:quasiparticle_GW_Z_fac}).
  
\subsection{Optical spectra with the Bethe-Salpeter equation}
The directionally averaged absorption cross section of a molecule is given by 
\begin{equation}
\begin{split}
\sigma(\omega) &= \frac{4\pi \omega}{3c} \sum_{m} \operatorname{Im}  \alpha_{m m}(\omega), \\
\end{split}
\label{eq:cross_section}
\end{equation}
where $ \alpha_{m m'}(\omega)$ is the dynamical dipole polarizability tensor given by
\begin{equation}
\begin{split}
\alpha_{m m'}(\omega) &= -\int  d^3r d^3r'  r_m \, \chi(\bm{r},\bm{r}', \omega)
\,  r'_{m'} \, .
\end{split}
\label{eq:alpha}
\end{equation}
The interacting density response function, or reducible polarizability, $\chi(\bm{r}, \bm{r}',\omega)$
is defined in the time domain as a functional derivative
of the density with respect to the change of the external potential:
$\chi(\mathbf 1,\mathbf 2)\equiv\frac{\delta \rho(\mathbf 1)}{\delta U(\mathbf 2)}$.
Numbered bold indices, $ \bm{i} = \{ \mathbf r_i, \sigma_i, t_i \}$,
refer to space, spin, and time coordinates, whereas plain numbered indices contain space and spin,
 $ i = \{ \mathbf r_i, \sigma_i \}$.  $\chi(\mathbf 1,\mathbf 2)$ is
a two-point quantity and it is directly connected to the non-interacting density
response $\chi^0(\mathbf 1,\mathbf 2)$ in RPA or in TDDFT with semi-local functionals \cite{Petersilka:1996}.
However, when the Hamiltonian becomes non-local in space (as in the case of 
TDHF, TDDFT with hybrid functionals or Hedin's $GW$ approximation) one must
first find the retarded four-point polarizability
$L(\mathbf 1,\mathbf 2,\mathbf 3,\mathbf 4)$, and then obtain the two-point one using
the relation $\chi(\mathbf 1,\mathbf 2) = L(\mathbf 1,\mathbf 1^+,\mathbf 2,\mathbf 2)$ (see appendix \ref{sec:appendix_derivation_BSE}). 
 
The four-point polarizability $L(\mathbf 1,\mathbf 2,\mathbf 3,\mathbf 4)$ satisfies 
the Bethe-Salpeter equation as derived in appendix \ref{sec:appendix_derivation_BSE}. 
In the frequency domain the BSE can be written
\begin{multline}
L(1,2,3,4\, |\, \omega) = L^0(1,2,3,4\, |\, \omega) \\
+ \int d(5678) L^0(1,2,5,6\, |\, \omega)K(5,6,7,8)L(7,8,3,4\, |\, \omega)
\end{multline}
with $L^0(1,2,3,4 \, |\, \omega)$ the non-interacting four-point polarizability and
\begin{equation}
\begin{split}
K(1,2,3,4)= v(1,3)\delta(1,2) \delta(3,4) - W(1,2)\delta(1,3)\delta(2,4) \, ,
\label{eq:BSE_kernel}
\end{split}
\end{equation}
the BSE kernel. Already here the approximation has been made that the screened interaction 
$W(1,2)$ is independent of the frequency. Introducing an orthonormal two-particle basis $| ij \rangle$ 
that has the representation 
$\langle 1,2 | ij \rangle = %\langle r_1 \sigma_1 r_2 \sigma_2 | ij \rangle= 
 \psi_i(1) \psi_j^*(2)$  %=\psi_i(r_1, \sigma_1) \psi_j^*(r_2 , \sigma_2)  $ 
in terms of the quasiparticle spin orbitals, we can expand $L$ as
\begin{equation}
\begin{split}
L(1,2,3,4 \, |\,  \omega) &=\sum_{ij,kl} \langle 1, 2 | ij \rangle L_{ij,kl}(\omega) \langle k l | 3, 4 \rangle \\
&= \sum_{ij,kl} \psi_i(1) \psi_j^*(2) L_{ij,kl}(\omega) \psi_k^*(3) \psi_l(4) \,,
\end{split}
\label{eq:L_expansion}
 \end{equation}
 with the matrix elements given by
 \begin{equation}
\begin{split}
L_{ij,kl} (\omega)&= \int d(1234) \psi_i^*(1) \psi_j(2) L(1,2,3,4\, |\,  \omega) \psi_k(3) \psi_l^*(4) \, .
\end{split}
 \end{equation}
$L^0$ is expanded similarly. This leads to the matrix equation
\begin{equation}
\begin{split}
L_{ij,kl}(\omega) = L^0_{ij,kl}(\omega)  + 
\sum_{i'j',k'l'} L^0_{ij, i'j'}(\omega) K_{i'j',k'l'}L_{k'l',kl}(\omega) \, .
\label{eq:BSE_matrix}
\end{split}
\end{equation}
Equation (\ref{eq:BSE_matrix}) has to be inverted for each frequency 
which is computationally cumbersome.
Fortunately, with certain approximations, it can be reformulated 
as an effective eigenvalue problem that only has to be solved once. 
To proceed with this we choose as our one-particle states the quasiparticle 
states in which the interacting Green's function $G$ is assumed to be diagonal. 
This leads to $L^0$ being 
diagonal in the two-particle  basis
\begin{equation}
\begin{split}
 L^0_{ij,kl} (\omega)  = \frac{\delta_{ik} \delta_{jl} 
 (f_i-f_j)}{\omega-(\epsilon_{j} -\epsilon_{i}) +\text{i}\gamma}.
  \end{split}
  \label{eq:L0-ph}
\end{equation}
where $f_i$ denotes the occupation number of spin orbital  $\psi_i$. We put the expression 
(\ref{eq:L0-ph}) in equation (\ref{eq:BSE_matrix}), 
rearrange terms and get after some algebra
\begin{equation}
\begin{split}
L_{ij,kl}(\omega) = \left [(\omega  +\text{i}\gamma)\delta_{i'k'}\delta_{j'l'} 
- H^\text{BSE}_{i'j',k'l'} \right ]^{-1}_{ij,kl} (f_k-f_l)  \, , 
\end{split}
\label{eq:L_inverse_H}
\end{equation}
where we introduced the frequency-independent BSE Hamiltonian
\begin{equation}
\begin{split}
H^\text{BSE} &= \sum_{ij, kl} | ij \rangle H^\text{BSE}_{ij,kl} \langle kl | \, ,\\
%H^\text{BSE}_{ij,kl}  &=& H^{0}_{ij,kl}+ (f_i-f_j) K_{ij,kl} \, .
H^\text{BSE}_{ij,kl}  &= (\epsilon_j -\epsilon_i)
\delta_{ik}\delta_{jl}+ (f_i-f_j) K_{ij,kl} \, .
\label{eq:BSE_Hamiltonian}
\end{split}
\end{equation}
The matrix $H^\text{BSE}$ is non-Hermitian. 
If we solve for its right eigenvectors and eigenvalues
\begin{equation}
H^\text{BSE} | 
\lambda \rangle =  \epsilon_{\lambda}  | \lambda\rangle \, ,	
\label{eq:BSE_eigenvalue}
\end{equation}
and define expansion coefficients of the eigenvectors in
terms of the the two-particle basis 
$A^{\lambda}_{ij} = \langle ij | \lambda \rangle$, 
we can obtain a spectral representation of the interacting polarizability as
\begin{equation}
\begin{split}
L_{ij,kl}(\omega) &=  \sum_{\lambda, \lambda'}
\frac{A^{\lambda}_{ij}  S^{-1}_{\lambda, \lambda'} A^{\lambda' *}_{kl} 
(f_k-f_l) }{\omega  -\epsilon_{\lambda} +\text{i}\gamma  }  \, .\end{split}
\label{eq:L_r}
\end{equation}
Here the overlap of the right eigenvectors $S_{\lambda, \lambda'} 
= \sum_{ij} A^{\lambda *}_{ij} A^{\lambda'}_{ij}$ appears because 
the eigenvectors of a non-Hermitian eigenvalue problem are
generally not orthogonal. 
Using equations (\ref{eq:alpha}), (\ref{eq:L_expansion}) and a resolution
of the identity in the quasiparticle product states, we can rewrite
$\alpha_{m m'}(\omega)$  in terms of $L$ as
\begin{equation}
\alpha_{m m'}(\omega) =\sum_{ijkl} 
D^{m *}_{ij} L_{ij,kl}(\omega) D^{m'}_{kl} \, ,
\label{eq:abs_spectra_non_TDA}
\end{equation}
with the transition dipoles
\begin{equation}
\begin{split}
D^{m}_{ij} &= \langle ij | D_m \rangle =  \int d(1) \psi^*_{i}(1) r_{m} \psi_{j}(1)\\
&= \delta_{x_i, x_j}\int d^3r \psi^*_{i}(\bm{r}) r_{m} \psi_{j}(\bm{r}) \, .
\end{split}
\end{equation}
Here $\psi_{i}(\bm{r}) $ is the spatial part of $\psi_{i}(1)$, and $x(\sigma)$ is the 
corresponding spin function. Here we denote the dipole operator as a ket, since 
in general a normal two-point operator $A$ can be expanded as 
$ A = \sum_{ij}A_{ij} | i \rangle \langle j | \equiv  \sum_{ij}| i j \rangle A_{ij} $. 
In the preceding analysis spin is explicit in the orbitals. 
However, $H^\text{BSE}$ is not diagonal in a spin orbital basis. 
If it is diagonalized 
in the spin indices (see appendix \ref{sec:appendix_spin}), one 
singlet and three triplet product functions result,  where the 
singlet one being the only one to have a non-vanishing transition 
dipole moment and so the one visible in the optical response. 
In the following we will suppress the spin indices and only 
work with the space quantities. Because of spin symmetry the 
coupling elements $K$ are modified with the factor $f^\text{s/t}$ 
being $2$ for the singlet and $0$ for the triplet    
\begin{equation}
\begin{split}
K_{ij,kl} &= f^\text{s/t}H^\text{ex}_{ij,kl} +H^\text{dir}_{ij,kl} \, ,\\
H^\text{ex}_{ij, kl} &=\int d^3r \,d^3r' \psi_i^*(\bm{r}) 
\psi_j(\bm{r}) v(\bm{r},\bm{r'}) \psi_k(\bm{r'}) \psi_l^*(\bm{r'})   \, ,\\
H^\text{dir}_{ij, kl} &=
-\int  d^3r \,d^3r' \psi_i^*(\bm{r}) \psi_k(\bm{r}) 
W(\bm{r},\bm{r'}) \psi_j(\bm{r'}) \psi_l^*(\bm{r'})  \, ,
\label{eq:kernel_integrals}
\end{split}
\end{equation}
and the transition dipoles for the singlet get an additional factor of $\sqrt{2}$ (see  appendix \ref{sec:appendix_spin})
\begin{equation}
\begin{split}
D^{m, singlet}_{ij} &= \sqrt{2} \int d^3r \psi^*_{i}(\bm{r}) r_{m} \psi_{j}(\bm{r}) \, .
\end{split}
\label{eq:transition_dipole_singlet}
\end{equation}
and  the triplet transition dipole is zero.
This means that the dynamic dipole polarizability  
effectively gets an additional factor of two for 
the singlet transition. Since we always consider the singlet 
for dipole transitions we can drop the "singlet" superscript and let 
$D^{m}_{ij}$ refer to equation \ref{eq:transition_dipole_singlet}.
An important simplification to the problem is that, due 
to the occupation factors, only particle-hole and 
hole-particle product states contribute to the polarizability 
(see appendix \ref{sec:appendix_spin}) 
and we can write the eigenfunctions of $H^\text{BSE}$ as
\begin{equation}
| \lambda \rangle = \sum_{vc}  | vc \rangle 
A^{ \lambda}_{vc}+ \sum_{vc}  |cv \rangle A^{\lambda}_{cv} \, .
\label{eq:BSE_eigenvector}
 \end{equation}
 Here and in the following the indices 
 $\{vv'\}$ will denote occupied (valence), 
 $\{ cc'\}$ empty (conduction, unoccupied)
 and $\{ ijkl\}$ general molecular orbitals. 
Projecting the eigenvalue equation (\ref{eq:BSE_eigenvalue}) 
from the left with $\langle vc |$ and $\langle cv |$ we
obtain a matrix equation with the following block structure 
\begin{equation}
\begin{split}
\left( \begin{array}{cc}
H^{0}_{vc, v'c'}  +K_{vc, v'c'} & K_{vc, c'v'}\\
- K_{cv, v'c'} & H^{0}_{cv, c'v'} -K_{cv, c'v'} \\
\end{array} \right) \left( \begin{array}{c}
A^{\lambda}_{v'c'}\\
A^{\lambda}_{c'v'}\\ 
\end{array} \right) = \epsilon_{\lambda} \left( \begin{array}{c}
A^{\lambda}_{vc}\\
A^{\lambda}_{cv}\\ 
\end{array} \right)  \, ,
\end{split}
\label{eq:BSE}
\end{equation}
where $H^{0}_{ij, kl} = (\epsilon_j -\epsilon_i)\delta_{ik}\delta_{jl}$. 
Using the symmetry properties of the BSE kernel 
$K_{ij,kl} = K^*_{ji, lk} =K^*_{kl,ij}$ and of the 
non-interacting Hamiltonian 
$H^0_{ij,kl} = -H^0_{ji,lk}$, we can also write 
\begin{equation}
\begin{split}
&H^\text{BSE}=\left( \begin{array}{cc}
H^{0}_{vc, v'c'}  +K_{vc, v'c'} & K_{vc, c'v'}\\
- K_{vc, c'v'}^* & -(H^{0}_{vc, v'c'}  +K_{vc, v'c'})^* \\
\end{array} \right) \,,
\end{split}
\label{eq:BSE_symm}
\end{equation}
The second form 
(\ref{eq:BSE_symm}) is useful 
because it leads to computational savings when explicitly 
setting up the matrix. 
 In the Tamm-Dancoff approximation the 
off-diagonal blocks in the $H^\text{BSE}$ (i.e. 
the couplings between hole-particle and 
particle-hole states) are set to zero.  
This leads to two
uncoupled Hermitian eigenvalue equations for  
$A^{\lambda}_{vc}$ and $A^{\lambda}_{cv}$. 
Due to the  symmetries displayed in equation 
(\ref{eq:BSE_symm}) we see that the eigenvalues of the two blocks are 
related as $\epsilon^{vc}_{\lambda} = -\epsilon^{cv}_{\lambda}$, 
and the eigenvectors  as $A^{\lambda}_{cv} = A^{\lambda,*}_{vc}$, 
where the superscript refers either to the $\{cv\}$ or the  $\{vc\}$-sector. 
Therefore, only one of the equations needs to be solved, for example the one for the
the $\{vc\}$-sector: $\sum_{v'c'} H^\text{res}_{vc, v'c'}   A^{\lambda}_{v'c'}   
= \epsilon_{\lambda} A^{\lambda}_{vc}$.
Using the fact that the eigenvectors are orthogonal for
a Hermitian problem, the non-zero blocks of the the four-point polarizability are
\begin{equation}
\begin{split}
L^\text{TDA}_{vc,v'c'}(\omega)=  \sum_{\lambda} \frac{A^{\lambda}_{vc} A^{\lambda *}_{v'c'}  }
{\omega  - \epsilon_{\lambda} +\text{i}\gamma }  \, , \\
 L^\text{TDA}_{cv,c'v'}(\omega)=  -\sum_{\lambda} \frac{A^{\lambda *}_{vc} A^{\lambda}_{v'c'}  }
{\omega  + \epsilon_{\lambda} +\text{i}\gamma }  \, . 
\end{split}
\end{equation}
The TDA is a widely used approximation that, in addition to the computational advantages,
often provide good agreement with experimental excitation energies for organic molecules \cite{Peach:2012, Peach:2011, Sharifzadeh:2013}.
At this point it is interesting to note the similarities of the 
BSE, TDDFT and time-dependent Hartree-Fock (TDHF). In TDDFT, although 
for semi-local functionals it is in principle sufficient to look at
the response of the density, one can more generally look at the
response of the density matrix as was done by Casida \cite{Casida:1995}.
The resulting equations are very similar to the BSE, with the only difference
that the $GW$ eigenvalues are replaced by KS eigenvalues, 
and that the direct term is replaced by a TDDFT
exchange-correlation kernel. 
For semi-local exchange-correlation functionals, and real orbitals, the
resulting eigenvalue problem can be reduced to a Hermitian problem of half
the size --- the preferred formulation of TDDFT in quantum chemistry.
However, when Hartree-Fock exchange is included (in hybrid functionals for
example)  the reduction to the Hermitian form does not simplify things quite as much, 
since one needs to take the square root of a full matrix which requires an additional diagonalization. 
The TDHF response equations have the same structure as
the BSE ones with Hartree-Fock eigenvalues and an unscreened direct
term. The Tamm-Dancoff approximation is also useful in TDDFT and TDHF.
For TDHF with TDA one recovers the configuration interaction singles (CIS)
equation.

To set up and diagonalize the BSE Hamiltonian (\ref{eq:BSE_symm}) is feasible only for systems with a few 
thousand of particle-hole pairs. For larger matrices an iterative procedure 
is essential both for memory and runtime requirements. In the following 
we describe how the the dynamical dipole 
polarizability tensor  (\ref{eq:abs_spectra_non_TDA}) can be computed with a  Lanczos-type iterative method.

\subsubsection{Continued fraction expression for the BSE polarizability}

Using equations (\ref{eq:L_inverse_H}) and (\ref{eq:abs_spectra_non_TDA}) 
 we can rewrite a matrix element of the dynamical dipole polarizability tensor (\ref{eq:alpha}) %(\ref{eq:cross_section}) 
in a form involving the resolvent of the
BSE Hamiltonian
\begin{equation}
\begin{split}
\alpha_{mm'}(\omega) &=
 -\sum_{ijkl} D^{m*}_{ij} L_{ij,kl} D^{m'}_{kl}\, \\
&=  -  \langle D_m | 
(\omega - H^\text{BSE} + i\gamma)^{-1}  | D_{m'}' \rangle ,
\end{split}
\label{eq:BSE_resolvent}
\end{equation}
where 
\begin{equation}
\begin{split}
|D_m \rangle &= \sum_{ij} | ij \rangle \langle ij | D_m \rangle \, ,\\
|D_{m'}' \rangle &= \sum_{ij} | ij \rangle (f_i-f_j) \langle ij | D_{m'} \rangle = 
\sum_{ij} | ij \rangle F_{ij,ij} \langle ij | D_{m'}' \rangle\, .
\end{split}
\label{eq:BSE_resolvent_dipoles}
\end{equation}
In the last equation $\langle ij | D_m \rangle$ refers to the singlet transition dipole in equation 
(\ref{eq:transition_dipole_singlet}), and we denote the occupation difference matrix by
\begin{equation}
F_{ij,kl} = (f_i -f_j)\delta_{ik} \delta_{jl}
\label{eq:F_occupations}
\end{equation}
In the Tamm-Dancoff approximation we only consider $\{vc\}$ states which means that the 
transition dipoles become
\begin{equation}
|D^{\text{TDA}}_m \rangle =  |D^{' \text{TDA}}_m \rangle = \sum_{vc} | vc \rangle \langle vc | D_m \rangle \, ,
\label{eq:BSE_resolvent_dipoles_TDA}
\end{equation}
(the TDA-superscript since it will be clear from the context if the TDA is used or not).  

An attractive method of dealing with resolvents
is the Haydock recursion scheme \cite{Haydock:1972}, where a diagonal 
matrix element of a resolvent  is efficiently computed from Lanczos 
recursion coefficients by means of continued fractions. 
Recently it has been shown that also non-diagonal 
matrix elements of the resolvent can be computed from the same Lanczos 
coefficients \cite{Rocca:2008, Gruening-Gonze:2011}.
Usually the continued fraction representation of the 
resolvent  is derived by determinant relations. Here we present an 
alternative derivation that only uses the power series expansion of 
the resolvent and the orthogonality  among the Lanczos vectors. The 
off-diagonal matrix elements come out naturally in this formulation, 
and it is straight-forwardly extendible to block Lanczos, two-sided 
Lanczos and pseudo-Hermitian Lanczos schemes. Our derivation also 
connects to the theory of relaxation functions, also known as the
Mori projection technique, first introduced to describe the Laplace 
transformed correlation function of  dynamical systems \cite{Mori:1965} 
and later  reformulated by Lee \cite{Lee:1982} in a form more 
closely related to the one we use here.   

We want to compute $ \langle i | (\omega -H)^{-1}  | j \rangle$ --- a 
general matrix element of the resolvent of the Hermitian operator $H$,
with the frequency $\omega$ in general a complex number. 
Let us define a frequency-dependent solution vector
\begin{equation}
| \tilde j(\omega)\rangle =  (\omega -H)^{-1}   | \tilde j \rangle \, ,
\label{eq:def_j_om}
\end{equation}
where $| \tilde j \rangle =|  j \rangle  / || j || $ is 
the normalized $| j \rangle$. The matrix element of 
the resolvent in terms of the solution vector (\ref{eq:def_j_om}) reads
\begin{equation}
  \langle   i | (\omega -H)^{-1}  | j \rangle =\langle i | 
  \tilde j(\omega)\rangle \cdot || j || \, .
\end{equation}
Now we generate a set of orthonormal Lanczos vectors $\{ |Êf_n \rangle \}$ with the starting 
state $ | f_0 \rangle=  | \tilde j \rangle$, 
using the standard recursion relations \cite{Saad:2003}
\begin{equation}
 b_{n+1} |  f_{n+1} \rangle  = H | f_{n} \rangle  - | 
 f_{n} \rangle a_n  -   |  f_{n-1} \rangle b_n \, ,  
\label{eq: lanczos_relation}
\end{equation}
with the real coefficients $a_n = \langle f_{n} | H | f_{n} \rangle$ and 
$b_n = \langle f_{n-1} | H | f_{n} \rangle$. 
Next we
expand  the solution vector $|\tilde j(\omega)\rangle$ in 
the Lanczos basis %$\{ |f_n \rangle \}$ 
\begin{equation}
 |  \tilde j(\omega)\rangle = \sum_n | f_n \rangle c_n(\omega) \, ,
\end{equation}
where the frequency dependent expansion 
coefficients $c_n(\omega)$ are given by projection onto the basis  
\begin{equation}
c_n(\omega) = \langle f_n  |  \tilde j(\omega)\rangle  \, .
\label{eq:c_n_def}
\end{equation}
The expansion coefficients $c_n(\omega)$ contain the information
necessary to compute the sought matrix 
elements of the resolvent. The diagonal matrix element 
is especially simple 
(remembering that $ | f_0 \rangle=  | \tilde j \rangle$)
\begin{equation}
\langle   j |(\omega -H)^{-1}  |j \rangle =  
\langle \tilde  j | \tilde j(\omega) \rangle \cdot || j ||^2 =   
c_{0}(\omega) \cdot || j ||^2   \, ,
\label{eq:mat_elem_diag}
\end{equation}
that is, only the zero-th coefficient $c_{0}(\omega)$ is needed.

In the original Haydock recursion scheme only diagonal matrix
 element were computed. For our purposes we also need the off-diagonal 
 elements, which can be computed using the higher expansion coefficients 
\begin{equation}
 \langle i |(\omega -H)^{-1}  | \tilde j \rangle = 
 \langle  i | \tilde j(\omega) \rangle = \sum_n \langle  i | f_n \rangle c_{n}(\omega) \, .
 \label{eq:mat_elem_nondiag}
\end{equation}

The 
projections $\langle  i | f_n \rangle$ of the vectors  $\langle  i |$ 
with the Lanczos basis can be computed and saved when the 
Lanczos vectors are available, thus avoiding the storage of more 
than the last two vectors. As we shall see, the coefficients $c_n(\omega)$ can be 
computed from continued fractions. An advantage of using continued fractions is that one can terminate 
them in a physically sensible way which can reduce the number of 
Lanczos vectors one has to explicitly compute. 
Projecting the Hermitian transpose of 
equation (\ref{eq: lanczos_relation}) onto the solution vector 
$|  \tilde j(\omega)\rangle $  
gives
\begin{equation}
\begin{split}
b_{n+1} \langle  f_{n+1} |  \tilde j(\omega)\rangle  = \langle  f_{n} | H |  \tilde j(\omega) \rangle - a_n  \langle f_{n} |  \tilde j(\omega) \rangle \\
 -b_n  \langle  f_{n-1}|  \tilde j(\omega)\rangle \, .
\end{split}
\label{eq:lanczos_coeffs_proto}
\end{equation}
Applying the operator $H$ onto the 
solution vector gives
\begin{equation}
H |  \tilde j(\omega)\rangle = \omega 
| \tilde j(\omega)\rangle - | \tilde j \rangle \, .
\label{eq:H_j_om}
\end{equation}
which follows directly from the definition of the inverse
\begin{equation}
 (\omega -H)(\omega -H)^{-1} =  \mathbbm{1}
%\label{eq:H_om_H}
\end{equation}
together with the definition of the solution vector 
$|  \tilde j(\omega)\rangle $ (\ref{eq:def_j_om}). %directly leads to equation (\ref{eq:H_j_om}).
Inserting equation (\ref{eq:H_j_om}) into equation 
(\ref{eq:lanczos_coeffs_proto}) we obtain a recursion relation 
for the expansion coefficients $c_{n}(\omega)$
\begin{equation}
b_{n+1} c_{n+1}(\omega)   = \omega c_{n}(\omega) - 
\delta_{n0} - a_n   c_{n}(\omega)- b_n c_{n-1}(\omega)\, .
\end{equation}
For $n=0$ the relation can be rearranged to give
\begin{equation}
c_{0}(\omega) = [\omega - a_0 - b_1 c_{1}(\omega) c^{-1}_{0}(\omega) ]^{-1}\, ,
\label{eq:c_0}
\end{equation}
while for $n>0$ we obtain 
\begin{equation}
c_{n}(\omega) c^{-1}_{n-1}(\omega)b_n^{-1} = 
[\omega - a_n - b_{n+1} c_{n+1}(\omega) c^{-1}_{n}(\omega) ]\, .
\label{eq:c_n}
\end{equation}

We now introduce the relaxation functions of order
$n$ $\varphi_n(\omega)$ \cite{Mori:1965, Lee:1982}
\begin{equation}
\begin{split}
\varphi_0(\omega) &= c_{0}(\omega)\, ,\\ 
\varphi_n(\omega) &= c_{n}(\omega) c^{-1}_{n-1}(\omega)b_n^{-1}, n>0 \, .
 \end{split}
\label{eq:phi_cont_frac}
\end{equation}
After inserting the expansion coefficients (\ref{eq:phi_cont_frac}) 
in equations (\ref{eq:c_0}), (\ref{eq:c_n}) we obtain 
the continued fraction relations familiar from the Haydock recursion scheme
\begin{equation}
\varphi_n(\omega) = [\omega - a_n - b_{n+1}^2 \varphi_{n+1}(\omega) ]^{-1}\, .
\label{eq:relax_fun_cont_frac}
\end{equation}

After the relaxation functions have been computed for a certain frequency, 
the expansion coefficients $c_n(\omega)$ can be recovered by inverting the 
relation (\ref{eq:phi_cont_frac})
\begin{equation}
\begin{split}
c_{n}(\omega) &= \varphi_{n}(\omega) b_n c_{n-1}(\omega) \\
& = \varphi_{n}(\omega) b_n \varphi_{n-1}(\omega) b_{n-1} \cdots \varphi_{1}(\omega) b_{1} \varphi_{0} (\omega)\, .
\label{eq:c_n_omega}
\end{split}
\end{equation}

In summary, first the coefficients $a_n$ and $b_n$, as 
well as the needed projections $\langle  i | f_n \rangle$  
are obtained from equation (\ref{eq: lanczos_relation}), 
then for each $\omega$ (adding a small positive imaginary part, 
as appropriate for the retarded response), the relaxation 
functions $\varphi_n(\omega)$ are computed from 
equations (\ref{eq:relax_fun_cont_frac}) 
using a properly chosen terminator. 
Then, the expansion coefficients $c_n(\omega)$  
are obtained from equation (\ref{eq:c_n_omega}). 
Finally, the matrix elements are computed from 
equations (\ref{eq:mat_elem_diag}) and (\ref{eq:mat_elem_nondiag}). 

\subsubsection{Iterative non-TDA BSE }
The full BSE Hamiltonian is non-Hermitian which means that the Lanczos 
procedure outlined above must be modified. A two-sided Lanczos 
procedure  where both left and right eigenvectors are generated 
in the recursive procedure can be used, although it suffers 
from instability issues due to the loss of orthogonality 
between the Lanczos vectors, often requiring explicit 
reorthogonalization \cite{Greenbaum:1997,Brezinski-Wuytack:2001}.
It also involves twice the number of applications of the Hamiltonian.
Recently, a pseudo Hermitian algorithm was published that exploits the 
structure of the BSE eigenproblem to convert it into a 
Hermitian problem in a special scalar product \cite{Gruening-Gonze:2011}. In 
this algorithm one avoids the extra multiplication 
of the Hamiltonian that is present in the two-sided scheme. Below we summarize the
pseudo-Hermitian algorithm in our notation.

An operator $A$  is pseudo-Hermitian \cite{Mostafadeh:2002}  with respect to the invertible Hermitian operator $\eta$, if
\begin{equation}
A = \eta^{-1} A^{\dagger} \eta .
\label{eq:pseudo_hermitian}
\end{equation}
This means that $\eta A$ is Hermitian,  or equivalently that $A$ is Hermitian under the scalar product  
$\langle \cdot | \cdot \rangle_{\eta} = \langle \cdot | \eta\cdot \rangle$, 
provided that the metric $\eta$ is positive definite so that the scalar 
product is well-defined. Furthermore, the eigenvalues of $A$ are real if it is pseudo-Hermitian with respect to an operator that can be written like $\eta = OO^{\dagger}$ with $O$ an invertible operator \cite{Mostafadeh:2002-2}, and such a factorization can always be found for a positive definite $\eta$.  If $A$ is a product of two Hermitian operators $A=BC$, then $A$ is 
pseudo-Hermitian with $B^{-1}$ and $C$, which can be checked using equation (\ref{eq:pseudo_hermitian}).
The BSE Hamiltonian $H^\text{BSE}$ given by 
equation (\ref{eq:BSE_Hamiltonian}) can be written in matrix form 
\begin{equation}
H^\text{BSE}=  H^{0} + F K \, , 
\end{equation}
with  $F$ given by equation (\ref{eq:F_occupations}).  %$F_{ij,kl} = (f_i -f_j)\delta_{ik} \delta_{jl}$. 
Since $F^2=I$ we can write 
\begin{equation}
\begin{split}
H^\text{BSE}=  F \bar H \, ,
\end{split}
\end{equation}
where
\begin{equation}
\begin{split}
\bar H = F H^{0} + K\, . 
\label{eq:H_BSE_bar}
\end{split}
\end{equation}
Since $FH^0$ is diagonal and real, and
$K_{ij,kl} = K^*_{kl,ij}$, it follows that $\bar H$ is Hermitian.
From the preceding discussion it is clear that 
$ H^\text{BSE}$ is pseudo-Hermitian with respect 
to $\eta=F^{-1}$ or $\eta=\bar H$.
Since $F$ is not positive definite it doesn't serve as
a metric for a scalar product. $\bar H$ however, should be positive
definite unless there exist singlet-triplet instabilities \cite{Bauernschmitt:1996, Peach:2012, Peach:2011}. Such instabilities do occur for molecules, and especially for triplet excitations $\bar H$ can lose its positive definiteness. This will make the pseudo-Hermitian algorithm fail. However, since in this case also direct diagonalization gives unphysical results one should not view this failure as a drawback of the method.  

Within the pseudo-Hermitian Lanczos 
scheme the same steps are followed as in the Hermitian case.  
The only difference is that the scalar product is changed form 
the ordinary $\langle \cdot | \cdot \rangle$ to $\langle \cdot | \bar H \cdot \rangle$, with the Lanczos vectors orthonormal in this product. This means that equation (\ref{eq: lanczos_relation}) stays the same, but the Lanczos coefficients are modified to $a_n = \langle f_{n} | \bar H  H^\text{BSE}  | f_{n} \rangle$ and 
$b_n = \langle f_{n-1} | \bar H H^\text{BSE}  | f_{n} \rangle$, which can be seen by multiplying equation (\ref{eq: lanczos_relation}) by $\bar H$ and using the orthogonality of the Lanczos vectors in the  $\langle \cdot | \bar H \cdot \rangle$ scalar product. To make the starting vector normalized, it is chosen as 
$ | f_0 \rangle = | \tilde j \rangle =  
| j \rangle  \langle j | \bar H  | j \rangle^{-1/2}$. 

Due to the metric introduced in our scalar product we 
effectively have right and 
left Lanczos vectors, related by 
$| f_n^L \rangle = \bar H| f_n^R \rangle$, 
and $| f_n^R \rangle = | f_n \rangle$, although only one set of vectors is
 necessary in the actual computation. The resolution of the identity in the Lanczos vectors is 
\begin{equation}
1 = \sum_n | f_n^R\rangle \langle f_n^L| = \sum_n | 
f_n^R\rangle \langle f_n^R | \bar 
H= \sum_n | f_n\rangle \langle f_n | \bar H,
\end{equation}
which means that the matrix element of the resolvent must be computed as
\begin{equation}
\begin{split}
 \langle i | ( \omega -H)^{-1}  | \tilde j \rangle &= 
 \sum_n \langle  i | f_n\rangle \langle f_n| 
 \bar H  | \tilde j(\omega) \rangle \\
 &= \sum_n \langle  i | f_n \rangle c_{n}^{\bar H}(\omega).
\end{split}
\end{equation}
Here $ c_{n}^{\bar H}(\omega) = \langle f_n| 
 \bar H  | \tilde j(\omega) \rangle $ replaces equation (\ref{eq:c_n_def}) --- the other 
 equations that are needed can be derived as in the Hermitian case, only
 replacing the scalar product. Here, even if we only want
a diagonal matrix element we have to sum over the projections of all the Lanczos 
vectors, because the starting (right) vector is not orthogonal (in the ordinary
scalar product) to the other Lanczos vectors.
\subsection{Implementation of the Bethe-Salpeter equation }

Having a general description of the BSE and of an iterative algorithm
for solving it, we will describe our implementation using local basis functions. 

\subsubsection{Non-iterative algorithm}

It is straightforward to compute the matrix in equation (\ref{eq:BSE_symm}) and 
diagonalize it to obtain the four-point polarizability from equations (\ref{eq:L_r}) 
and  (\ref{eq:abs_spectra_non_TDA}). The matrix elements of the kernel $K$ are 
computed using equation (\ref{eq:kernel_integrals}). 
The construction of the matrix requires $O(N^5)$
operations ($N$ being the number of atoms) 
and $O(N^4)$ memory for storage. Solving the resulting eigenvalue 
problem using standard diagonalization techniques gives an even more 
prohibitive scaling of $O(N^6)$ with the number of atoms. A way to avoid 
this excessive scaling is to limit the number of electron-hole pairs 
that are included in the calculation. However, the energy range covered 
by a constant number of pairs decreases with increasing system size, 
leading to a deteriorated description of the spectrum.
In practice, the limit where explicit diagonalization is feasible is reached for a few tens of atoms: for
larger systems iterative schemes are more efficient. Nevertheless,
for small systems and for testing purposes straightforward diagonalization
is a simple and useful alternative. Using our localized product basis set $\{F_{\mu}(r) \}$,
the exchange and direct terms in equation (\ref{eq:kernel_integrals}) take the following form 
\begin{equation}
\begin{split}
&H^\text{ex}_{ij,kl} = 
\boxed{\sum_{\mu} \tilde V^{ij*}_{\mu} 
\boxed{\sum_{\nu} \tilde V^{kl}_{\nu} v_{\mu \nu}} } \, ,\\
&H^\text{dir}_{ij,kl}= 
\boxed{ -\sum_{\mu} \tilde V^{ik*}_{\mu} 
\boxed{ \sum_{\nu} \tilde V^{jl}_{\nu} W_{\mu \nu} }}\, ,\\
\end{split}
\label{eq:coulomb_integrals_domprod} 
\end{equation}
where the bare and screened Coulomb matrix elements in the local product basis are
\begin{equation}
\begin{split}
v_{\mu\nu} &= \int  d^3r \, d^3r'  F^*_{\mu}(\bm{r}) v(\bm{r},\bm{ r'}) F_{\nu}(\bm{r'}) \, ,\\   
W_{\mu\nu} &= \int d^3r \, d^3r' F^*_{\mu}(\bm{r}) W(\bm{r},\bm{ r'}, \omega=0) F_{\nu}(\bm{r'}) \, . 
\end{split}
\label{eq:coulomb_mat_elem_domiprod} 
\end{equation}
The expansion coefficient $\tilde V^{ij}_{\mu}$ of a product of 
two quasiparticle states is given by 
\begin{equation}
\tilde V^{ij}_{\mu} = \boxed{\sum_{a} X_{ia} \boxed{\sum_{b} V^{ab}_{\mu} X^{*}_{jb}}} \, ,
\label{eq:vertex_ij}
\end{equation}
where the expansion coefficients $V^{ab}_{\mu}$ are those appearing in equation (\ref{eq:ao_vertex}).
Unlike the local product coefficients $V^{ab}_{\mu}$, the eigenstate product coefficients $\tilde V^{ij}_{\mu}$ are not sparse
and the equations (\ref{eq:coulomb_integrals_domprod}) 
will scale like $O(N^5)$ if the loops are ordered in the proper 
way as shown by the boxes (equation (\ref{eq:vertex_ij}) costs $O(N^4)$ operations). 
The singlet transition dipoles can also be 
calculated from the product functions
\begin{equation}
D_{ij}^{m} = \sqrt{2} \sum_{\mu} \tilde V_{\mu}^{ij} D_{\mu}^{m} \, ,
\end{equation}
where the dipole moments in the local product basis are $ D_{\mu}^m = \int d^3r F^*_{\mu}(\bm{ r}) r_m $.
\subsubsection{Iterative computation of the BSE }

Let us first look at the TDA 
which is simpler than the full BSE.
Because only the $\{vc \}$-sector needs 
to be solved, the eigenvalue problem is Hermitian. Moreover,
because $|D_m \rangle = |D_m' \rangle$ in equation (\ref{eq:BSE_resolvent_dipoles_TDA}), we only need to calculate a 
diagonal matrix element of the resolvent to get the diagonal dynamical dipole polarizability
\begin{equation}
\begin{split}
\alpha_{mm}(\omega)
&=  - \langle \tilde D_m | (\omega - H^\text{BSE} + \text{i}\gamma)^{-1}   | \tilde D_m \rangle \cdot  ||D_m||^2 \\
&= -c_0(\omega) \cdot ||D_m||^2.
\end{split}
\end{equation}

Here $| \tilde D_m\rangle = |D_m\rangle / ||D_m||$  in equation (\ref{eq:BSE_resolvent_dipoles}) is used as the starting vector in the Lanczos recursion. 
The dynamical dipole polarizability can directly  be written as a continued fraction 
using equations (\ref{eq:phi_cont_frac}) and (\ref{eq:relax_fun_cont_frac}): 
\begin{equation}
\begin{split}
\alpha_{mm}(\omega) &=  -\cfrac{  ||D_m||^2}{\omega +i\gamma -a_0 - \cfrac{b_1^2}{ \omega + i\gamma -a_1 
-\cfrac{b_2^2}{\cdots}}} \, .
\end{split}
\end{equation}
The Lanczos procedure for TDA is
\begin{empheq}[box=\fbox]{align}
& | f_{-1} \rangle = 0, \nonumber \\
& | \tilde f_{0} \rangle = |\tilde D_m \rangle, \nonumber \\
%&& b_{n+1} | f_{n+1} \rangle = H |f_n \rangle - a_n |f_n\rangle  - b_n |f_{n-1} \rangle \\
&  | \tilde f_{n+1} \rangle = H^{BSE} |f_n \rangle - a_n |f_n\rangle  -
 b_n |f_{n-1} \rangle, \nonumber \\
&b_{n+1} = \langle \tilde f_{n+1}| \tilde f_{n+1} \rangle^{1/2},  \nonumber\\
%\langle f_{n-1} | H |f_n \rangle = \langle f_n |f_n \rangle
& | f_{n+1} \rangle = | \tilde f_{n+1} \rangle / b_{n+1}, \nonumber \\
&a_n = \langle f_n | H^{BSE}  |f_n \rangle \, ,
\end{empheq}
where first a non-normalized vector $| \tilde f_{n+1} \rangle$ is computed and the $b^2_{n+1}$ coefficient is computed from its norm. The most time-consuming step in computing the Lanczos coefficients is 
the application of the Hamiltonian to a vector.  Generally, we express the Lanczos vector in the $| vc \rangle$, $| cv \rangle$ basis, similarly to the BSE eigenvectors in equation (\ref{eq:BSE_eigenvector}) 
 \begin{equation}
| f_n \rangle = \sum_{vc}  | vc \rangle 
f^{vc}_{n} +  |cv \rangle f^{cv}_{n} \, ,
\label{eq:Lanczos_cv_vc_basis}
 \end{equation}
with the expansion coefficients
 \begin{equation}
 f^{vc}_{n}  =  \langle vc | f_n \rangle \, , \qquad f^{cv}_{n}  =  \langle cv | f_n \rangle .
 \end{equation}
In the TDA we only make use of the $| vc \rangle$ functions. We want to find the expansion coefficients of the vector
resulting from the application of the Hamiltonian, 
that is  $ \langle vc | H^{BSE}  |f_n \rangle$ 
The action of non-interacting part $H^0$ is evaluated in $O(N^2)$ operations
\begin{equation}
\langle vc | H^{0} |f_n \rangle = (\epsilon_c -\epsilon_v)
 f_{n}^{vc} .  %,  k=unocc, l =occ 
\end{equation}
To exploit 
the sparsity in the kernels $H^\text{ex}$ and $H^\text{dir}$, we also make use of an atomic orbital product basis $| ab\rangle$ with real space representation 
$\langle \bm{r} \bm{r'} | ab\rangle = f_a(\bm{r})f_b^*(\bm{r'}) $. Using the expansion of the quasiparticle states in AOs, equation (\ref{eq:MO_expansion}) we have 
\begin{equation}
 | ij\rangle = \sum_{ab}  | ab\rangle X_{ia} X_{jb}^* \,  ,
\end{equation}
which allows us to rewrite the kernel as
\begin{equation}
K_{ij,kl} = \sum_{ab,a'b'} X_{ia}^* X_{jb} K_{ab, a'b'} X_{ka'} X^*_{lb'}.
\end{equation}
with the matrix elements of the kernel expressed in the AO basis
\begin{equation}
\begin{split}
K_{ab,a'b'} &= f^\text{s/t}H^\text{ex}_{ab,a'b'} +H^\text{dir}_{ab,a'b'} \, ,\\
H^\text{ex}_{ab, a'b'} &=\int d^3r \,  d^3r' f_a^*(\bm{r}) 
f_b(\bm{r}) v(\bm{r},\bm{r'}) f_{a'}(\bm{r'}) f_{b'}^*((\bm{r'})   \, ,\\
H^\text{dir}_{ab,a'b'} &=
-\int d^3r \,  d^3r' f_a^*(\bm{r}) f_{a'}(\bm{r}) 
W(\bm{r},\bm{r'}) f_{b}(\bm{r'}) f_{b'}^*(\bm{r'})  \, .
\label{eq:kernel_integrals_AO}
\end{split}
\end{equation}
The application of $K$ to a Lanczos vector becomes
\begin{equation}
 \langle vc | K |f_n \rangle= 
\sum_{ab} X_{va}^* X_{cb} \sum_{a'b'} K_{ab,a'b'} 
\sum_{v'c'} X_{v'a'} X^*_{c' b'} f_n^{v'c'}. 
\end{equation}
The operation  is separated in three steps: first the 
coefficient vector is transformed from the eigenstate basis to the local basis
\begin{equation}
f_n^{ab} =  \boxed{ \sum_{v} X_{va} \boxed{ \sum_c X^*_{cb} f_n^{vc}}} \, ,
\end{equation}
then the kernel $K$ is applied in the local basis
\begin{equation}
{f'}_{n}^{ab} =  \sum_{a'b'} K_{ab,a'b'}  f_{n}^{a'b'}\, ,
\end{equation}
and finally the coefficient vector is back transformed to the eigenstate basis
\begin{equation}
 \langle vc | K |f_n \rangle=  \boxed{ \sum_{a} X_{va}^* \boxed{\sum_{b} X_{cb} {f'}_{n}^{ab}}} \, .
\end{equation}
The transform and back-transform  
can be done in $O(N^3)$ operations since they consist of 
matrix-matrix multiplications which are done sequentially,
as shown by the boxes. The application of the kernel 
$K_{ab,a'b'}$ would generally take $O(N^4)$ operations, but due to sparsity it
takes actually $O(N^2)$ operations. $H^{ex}$ is expressed in the 
$\{F_{\mu}(\bm{r})\}$ basis as
\begin{equation}
\begin{split}
H^\text{ex}_{ab,a'b'}    
&=  \sum_{\mu, \nu} V^{* ab}_{\mu} v_{\mu \nu} V^{a'b'}_{\nu}.
\end{split}
\end{equation}
and the action on the coefficients becomes 
\begin{equation}
\sum_{a'b'} H^\text{ex}_{ab,a'b'} \, f_{n}^{a'b'} =
 \sum_{\mu \ni a, b} \sum_{\nu} \sum_{ a', b' \in \nu}  V^{ab*}_{\mu} v_{\mu \nu} V^{a'b'}_{\nu} \, c_{'n}^{a'b'}.
 \label{eq:action_ex}
\end{equation}
For the direct term we similarly get
\begin{equation}
\begin{split}
H^\text{dir}_{ab,a'b'} = - \sum_{\mu, \nu} V^{* aa'}_{\mu} W_{\mu \nu} V^{bb'}_{\nu},
\end{split}
\end{equation}
and the action of the coefficients are
\begin{equation}
\sum_{a'b'} H^\text{dir}_{ab,a'b'} \, f_{n}^{a'b'} = 
\sum_{a',b'} \sum_{\mu \ni a, a'} \sum_{\nu \ni b, b'}  
V^{*aa'}_{\mu} W_{\mu \nu} V^{bb'}_{\nu} \, f^{n}_{a'b'}.
\label{eq:action_dir}
\end{equation}
The Coulomb matrix elements $v_{\mu \nu}$ and $W_{\mu \nu}$ are given by equation (\ref{eq:coulomb_mat_elem_domiprod}).
Because by construction the matrix of product coefficients $V^{ab}_{\mu}$ 
is sparse, a fixed number of atomic orbitals 
couple for each $\mu$ or $\nu$ and the operations in equations (\ref{eq:action_ex}) 
and (\ref{eq:action_dir}) scale 
asymptotically as $O(N^2)$. 

For the solution of the full BSE problem we use the pseudo-Hermitian 
Lanczos scheme with the scalar product $\langle\cdot  | \bar H \cdot \rangle$ 
as explained in the previous section.  
A matrix element of the dynamical dipole polarizability computed with the iterative algorithm is given by
\begin{equation}
\begin{split}
\alpha_{mm'}(\omega)&=  -\langle D_m | 
(\omega - H^{\text{BSE}} + \text{i}\gamma)^{-1}  | D_{m'}' \rangle \\
&= - \sum_n \langle D_m  | f_n \rangle c_{n}^{\bar H}(\omega) 
\langle D_{m'}' | \bar H | D_{m'}' \rangle^{-1/2},
\end{split}
\end{equation}
where the coefficients $c_{n}^{\bar H}(\omega)$ are 
computed from the continued fractions $\varphi_n(\omega)$ 
as given by equations (\ref{eq:relax_fun_cont_frac}) and (\ref{eq:c_n_omega}).
Note that since we are already computing off-diagonal matrix elements there is
little extra cost to obtain the full dynamical dipole polarizability tensor, and not just the diagonal
matrix elements as is usually done in the TDA case.
The Lanczos procedure in the pseudo-Hermitian case is
\begin{empheq}[box=\fbox]{align}
& | f_{-1} \rangle = 0,\nonumber \\
& | \tilde f_{0} \rangle = |\tilde D_{m'}' \rangle,\nonumber \\
& | \tilde f_{0} '  \rangle = \bar H|\tilde f_{0} \rangle, \nonumber \\
& b_0 = \langle \tilde f_{0}' |  \tilde f_{0} \rangle^{1/2}, \nonumber \\
& | f_{0} \rangle = | \tilde f_{0} \rangle / b_0,\nonumber  \\
& | f_{0}' \rangle = | \tilde f_{0}' \rangle / b_0, \nonumber \\
& \nonumber \\
& a_n =  \langle f_n' | F  |f_n' \rangle,\nonumber \\
&  | \tilde f_{n+1} \rangle = F   |f_n' \rangle - a_n |f_n\rangle  - b_n |f_{n-1} \rangle, \nonumber \\
&  | \tilde f_{n+1}' \rangle = \bar H |\tilde f_{n+1} \rangle, \nonumber \\
& b_{n+1} = \langle \tilde f_{n+1}' |  \tilde f_{n+1} \rangle^{1/2}, \nonumber \\
& | f_{n+1} \rangle = | \tilde f_{n+1} \rangle / b_{n+1}, \nonumber \\
& | f_{n+1}' \rangle = | \tilde f_{n+1}' \rangle / b_{n+1}. 
\end{empheq}
In this scheme the intermediate vector $| {f'}_n \rangle$ is saved between iterations in order to minimize the number of applications of the Hamiltonian.  To perform a Lanczos iteration we need to apply $\bar H$ given by 
equation (\ref{eq:H_BSE_bar}) to some vector 
$| f_n \rangle$, now containing both particle-hole 
and hole-particle amplitudes. This in done much in 
the same way as in the TDA case. The term $FH^0$
is diagonal in the eigenstate basis and becomes
\begin{equation}
\begin{split}
\langle vc | F H^0 | f_n \rangle =& 
\sum_{v'c'} (FH^0)_{vc,v'c'} f_n^{v'c'} = 
(\epsilon_v -\epsilon_c) f_n^{vc} \, ,\\
\langle cv | F H^0 | f_n \rangle =& 
\sum_{c'v'} (FH^0)_{cv,c'v'} f_n^{c'v'} = 
(\epsilon_v -\epsilon_c) f_n^{cv}  \, .
\end{split}
\end{equation}
For the coupling matrix elements we transform to the atomic basis as in the TDA case, with the exception that we need to make use of both $| vc \rangle$ and $| cv \rangle$ vectors. The transformation is done for both sets of vectors as shown below
\begin{equation}
f_n^{ab} =  \boxed{ \sum_{c} X_{ca} \boxed{ \sum_v X^*_{vb} (f_n^{vc} +f_n^{cv}) }} \, .
\label{eq:aux_vectors_ao}
\end{equation}
After the auxiliary vector in equation (\ref{eq:aux_vectors_ao}) has been computed, the exchange and direct terms are applied in exactly 
the same way as in the TDA case, and after that the coefficients are back-transformed as
\begin{equation}
\begin{split}
{f'}_{n}^{vc} =  \boxed{ \sum_{a} X_{va}^* \boxed{\sum_{b} X_{cb}  {f'}_{n}^{ab}}} \, \\ 
{f'}_{n}^{cv} =  \boxed{ \sum_{a} X_{ca}^* \boxed{\sum_{b} X_{vb}  {f'}_{n}^{ab}}} \, . 
\label{eq:aux_vectors_mo}
\end{split}
\end{equation}
Finally we need to apply $F$, which is easily done considering its definition (\ref{eq:F_occupations})
\begin{equation}
\begin{split}
\langle vc |  F| f_n \rangle =& \sum_{v'c'} 
F_{vc,v'c'} f_n^{v'c'}  = f_n^{vc} \, , \\
\langle cv |  F| f_n \rangle =& \sum_{v'c'} F_{cv,c'v'} f_n^{c'v'}  = 
-f_n^{cv} \, .
\end{split}
\end{equation}
To conclude, we have shown that the application of the 
Hamiltonian onto a particle-hole state takes $O(N^3)$ operations, 
both when using the Tamm-Dancoff approximation and solving the full BSE. 
If we use the continued fraction method with a given broadening we 
can assume that the number of Lanczos coefficients will be 
independent \cite{Shirley:1999} of the number of atoms.
This then leads to an overall  $O(N^3)$ complexity scaling of the algorithm.

\section{Test calculations}

\subsection{Simple cases: Na$_2$ and CH$_4$}

\begin{figure}[t!]

\includegraphics[width=1.0\columnwidth, angle=0]{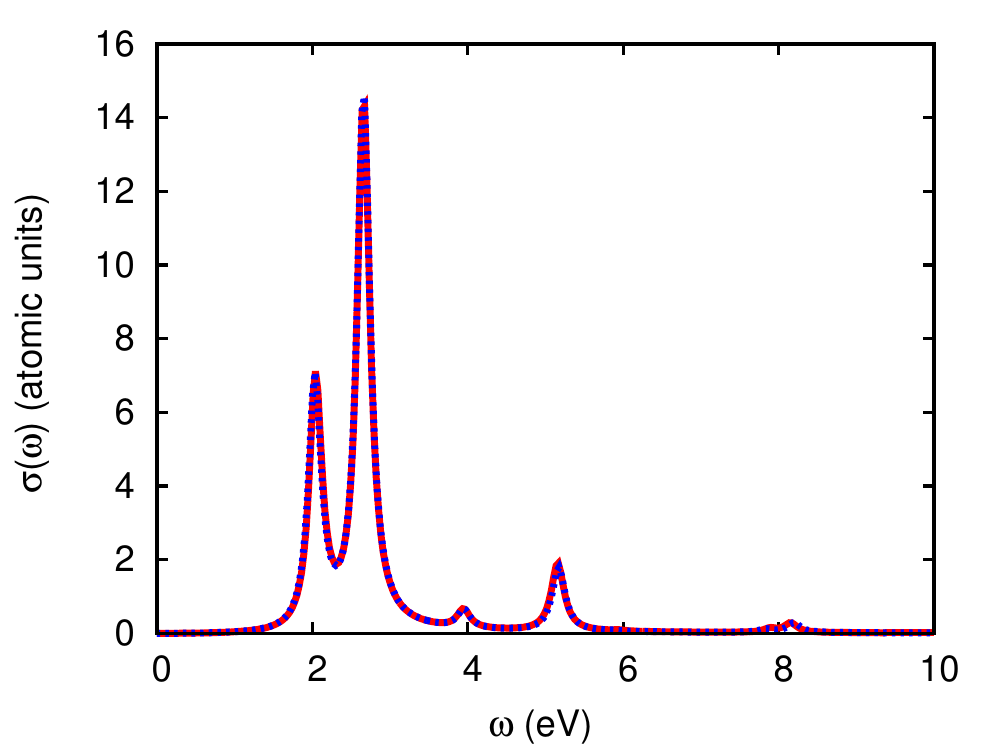}

\caption{Comparson of the absorption cross section
of the sodium dimer between our method (red full line) and 
MOLGW (blue dotted line). A Lorenzian broadening with FWHM of 0.2 eV was used in both cases. }
\label{fig:Na2_molgw}
\end{figure}

As a first test of the implementation we look at two simple 
test systems for which we can make accurate comparisons to other 
codes. The sodium dimer is simple in that it has only one 
valence orbital (filled with two electrons) which makes the 
spectrum dominated by single transitions.
We computed the $G_0W_0$/BSE for this system starting from an all-electron HF  
ground state calculation performed with our code, using the cc-pVDZ Gaussian basis set treated as 
numerical atomic orbitals. The $G_0W_0$ calculation was performed using 
fully frequency-dependent self energy in the range of the
valence and semi-core states while the 1s core orbitals were treated with HF exchange only. 
The quasiparticle energies were computed using the standard 
first order expansion of the $\operatorname{Re}\Sigma_{ii}(\omega)$ around the initial 
HF eigenvalue $\epsilon_i$ according to equation (\ref{eq:quasiparticle_GW_Z_fac}). %As described in \cite{Ljungberg_BSE_SS:2015} 
This procedure is less accurate than solving for the quasiparticle energy 
graphically, but here we are more interested in a comparison rather than a 
fully converged result. The BSE was solved by direct diagonalization. For 
comparison we use the MOLGW code by Bruneval \cite{Bruneval_molgw:2012,
Bruneval:2013,Bruneval_BSE:2015} 
where we as far as possible use the same parameters as in our code.  
In table \ref{tab:Na2_GW_BSE} we compare the first ionization potential (IP) 
and electron affinity (EA) as well 
as the position of the first BSE transition obtained with the two codes. 
The agreement is excellent. Furthermore, the computed cross sections 
for both TDA and full BSE match very well even for higher transitions ---
the obtained optical spectra lie on 
top of each other, as can be seen in the figure \ref{fig:Na2_molgw}.
In the calculation of absorption cross section a Lorentzian broadening
of 0.2 eV was used.

\begin{table}[htdp]
\caption{Comparison of calculated energies
obtained with our code and MOLGW code \cite{Bruneval:2013}
for the sodium dimer. In both calculations the cc-pVDZ basis set 
is used. Energies are given in units of eV.}
\begin{center}
\begin{tabular}{lcc}
\hline\hline
& This work & MOLGW\\
\hline
 IP (HF) &  4.53  & 4.54 \\
 %\hline
EA (HF)  & -0.14  &   -0.13  \\
%\hline
Gap (HF)  &  4.68 &  4.67 \\
%\hline
 IP ($G_0W_0$) &  4.88  & 4.88 \\
 %\hline
EA ($G_0W_0$)  &  0.17  & 0.18\\
%\hline
Gap ($G_0 W_0$)  & 4.71 &  4.70\\
%\hline
 BSE (TDA), singlet &  2.29   &   2.29 \\
% \hline
% \hline
BSE (full), singlet  &  2.04 &   2.03\\
% \hline
\hline\hline
\end{tabular}
\end{center}
\label{tab:Na2_GW_BSE}
\end{table}%

\begin{figure}[t!]

\includegraphics[width=1.0\columnwidth, angle=0]{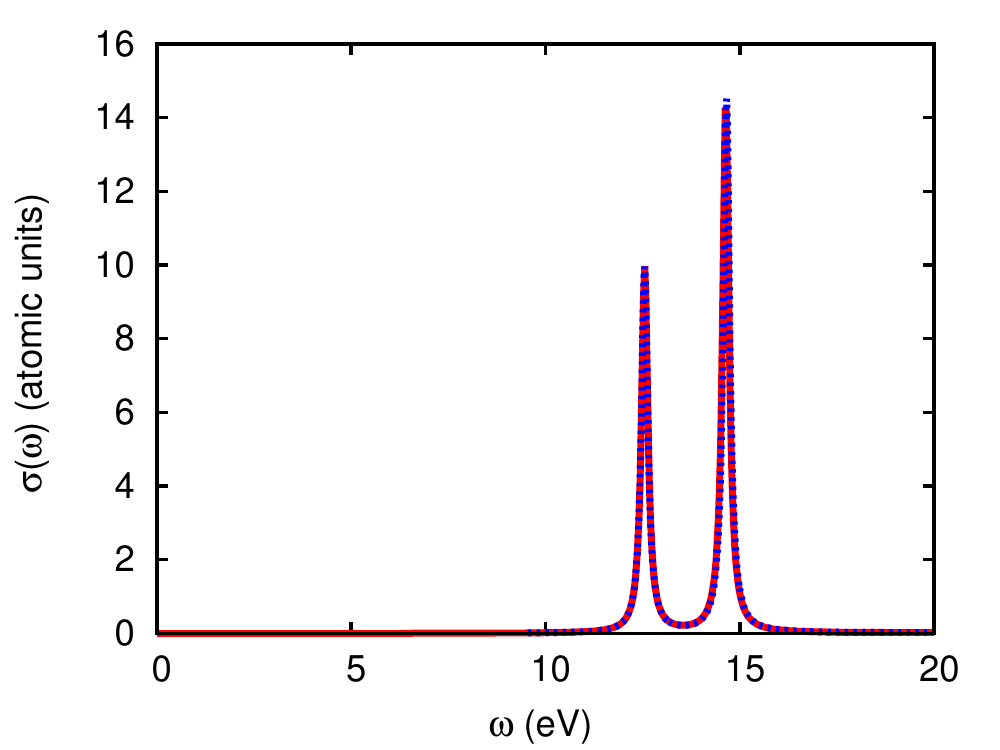}

\caption{Comparson of the absorption cross section
of CH$_4$  between our method (red full line) and 
MOLGW (blue dotted line). A Lorenzian broadening with FWHM of 0.2 eV was used in both cases.}
\label{fig:CH4_molgw}
\end{figure}

As a second example we chose to investigate the methane molecule, CH$_4$. Similarly to the 
case of sodium dimer, we have chosen cc-pVDZ basis in both calculations,
0.2 eV Lorenzian broadening and also followed as much as possible the 
same procedure to extract $G_0W_0$ eigenvalues. 
In table \ref{tab:CH4_GW_BSE} we make the same comparison as 
in previous example, with the same excellent agreement
for $GW$ energies, first optical excitation in TDA and
full BSE. The computed spectra are also in perfect agreement, 
as can be seen in figure  \ref{fig:CH4_molgw}.

\begin{table}[htdp]
\caption{Comparison of calculated energies
obtained with our method and the MOLGW code \cite{Bruneval:2013}
for methane. In both calculations the cc-pVDZ basis set 
is used. Energies are given in units of eV.}
\begin{center}
\begin{tabular}{lcc}
\hline\hline
& This work & MOLGW\\
\hline
 IP (HF) &  14.76  &  14.77\\
 %\hline
EA (HF)  & -5.26 & -5.25    \\
%\hline
Gap (HF) & 20.02   & 20.02 \\
%\hline
 IP ($G_0W_0$) &  14.40  &  14.41 \\
 %\hline
EA ($G_0W_0$)  & -4.81  & -4.81\\
%\hline  
Gap ($G_0 W_0$)  & 19.22 & 19.22 \\
%\hline
 BSE (TDA), singlet &  12.58  &   12.59 \\
% \hline
% \hline
BSE (full), singlet  &  12.55 &   12.55   \\
% \hline
\hline \hline
\end{tabular}
\end{center}
\label{tab:CH4_GW_BSE}	
\end{table}%

\subsection{Iterative method versus diagonalization}

Confident that our BSE matrix is set up correctly we now turn to the iterative 
method. As a more suitable test case we chose the benzene molecule that is small
enough for direct diagonalization (with a moderately large basis set) while still
having many transitions that contribute to the spectrum. The ground state 
calculation was done with the SIESTA code \cite{SIESTA} using the PBE functional and a  
DZP basis set, using an energy shift of 3 meV. Although this basis set is not fully converged for $GW$ quasiparticle energies and 
optical properties it gives reasonable results for the IP (8.85 eV) and EA (-1.34 eV) compared to 
earlier obtained results \cite{Foerster:2011}, and to experimental values \cite{NIST_database}. 
The first visible optical transition in our calculations occurs at 6.95 eV for the TDA and 6.18 eV for the full BSE, 
compared to the experimental 
value of 6.92 eV (extracted from the experiment shown in ref. \onlinecite{Tiago:2005}). We note that the effect of introducing the
TDA here is quite large.
A detailed account of the convergence properties of quasiparticle energies and BSE spectra for this system, as well as for larger organic molecules, 
will appear in a forthcoming publication \cite{Ljungberg_BSE_SS:2015}.  For the evaluation of the iterative method, the parameters we choose here are fully sufficient. 

\begin{figure}[t!]

\includegraphics[width=1.0\columnwidth, angle=0]{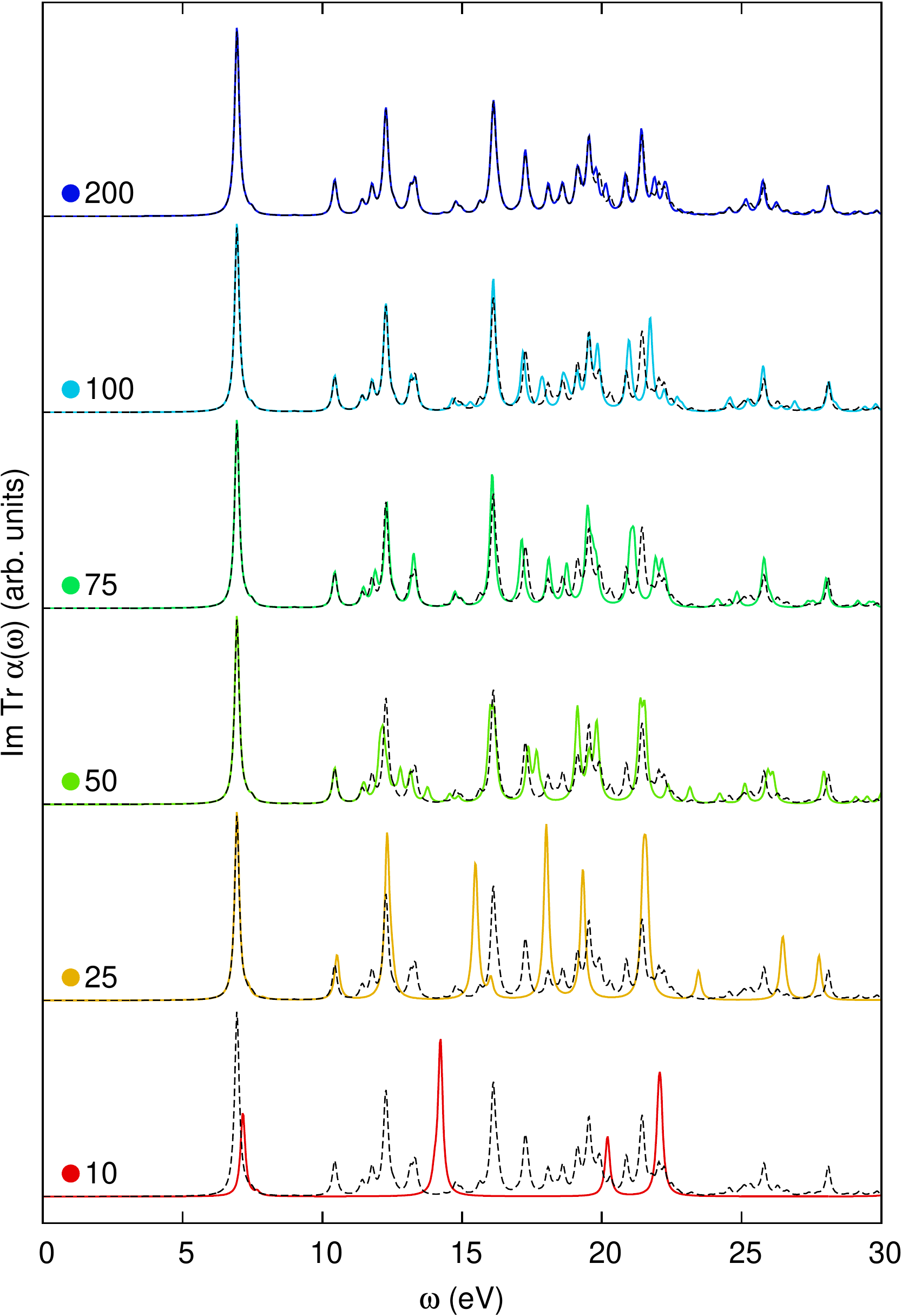}

\caption{The convergence of the trace of the TDA polarizability with the number of
recursion vectors for benzene. The results obtained with 10, 25, 
50, 100 and 200 iterations are compared to direct diagonalization 
of the full BSE Hamiltonian (dashed lines).}
\label{fig:BSE_recursion_conv_TDA}
\end{figure}

\begin{figure}[t!]

\includegraphics[width=1.0\columnwidth, angle=0]{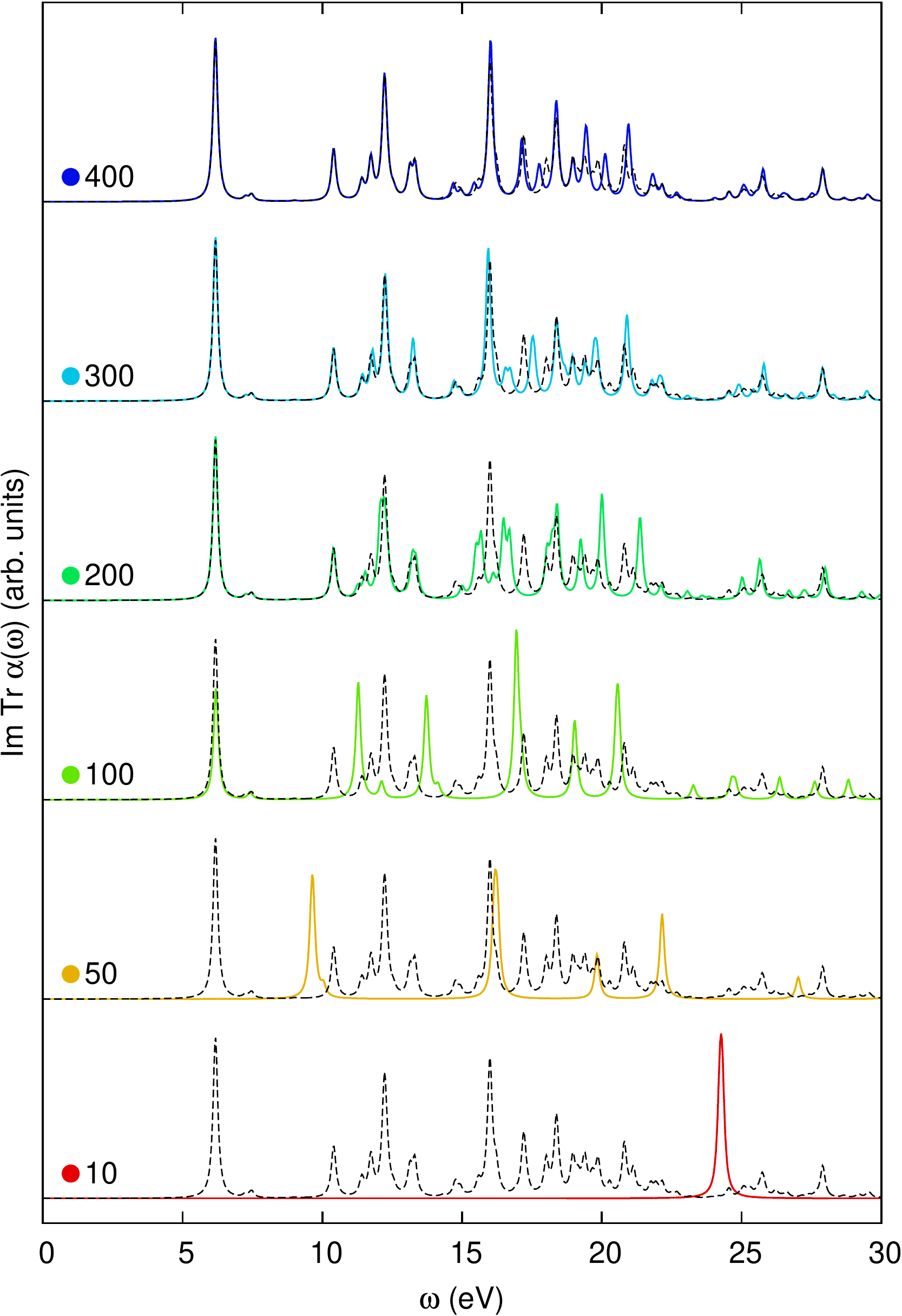}

 \caption{The convergence of the trace of the full BSE polarizability with the number of
recursion vectors for benzene. The results obtained with 10, 50, 
100, 200, 300 and 400 iterations are compared to direct diagonalization 
of the BSE Hamiltonian.}
\label{fig:BSE_recursion_conv_nonTDA}
\end{figure}

In figures \ref{fig:BSE_recursion_conv_TDA} and 
\ref{fig:BSE_recursion_conv_nonTDA}, we show the comparison of the iterative method 
for TDA and non-TDA to direct diagonalization.  A simple truncation of the continued 
fraction is used here. We see that the converged iterative spectrum is obtained with 
around 200 recursion coefficients for TDA and around 400 for the non-TDA spectrum 
for this broadening. % (FWHM  $=0.2$ eV). 
Note that the full particle-hole space has a dimension of 1400 
for TDA and 2800 for the full BSE.

Next we look at different terminators of the continued fraction. The last relaxation function in equation (\ref{eq:relax_fun_cont_frac}) is assumed to  satisfy 
\begin{equation}
\begin{split}
 \varphi_{n-1}(\omega) = [\omega -a_{n-1} -b^2_{n} \varphi_{T}(\omega) ]^{-1}
\end{split}
\end{equation}
where $\varphi_{T}(\omega)$ the terminator function. The simplest terminator is obtained by truncation, which means that the remaining coefficients that are not explicitly computed are set to zero. This gives  $\varphi_{T}(\omega) =1/\omega$, and corresponds to 
a representation of the dynamical dipole polarizability as a sum of delta functions. However, often a more suitable terminator can be found by extrapolating the remaining coefficients according to some physical model suited to the system of study.  In the first model we consider, the dynamical dipole polarizability is assumed to be a continuous distribution without gap,
centered at $a$ and 
with width $2E_W$.  In this case the  $a_n$ coefficients should converge to $a$, 
and $b_n$ should converge to $b=E_W/2$ \cite{Haydock:1972, Turchi:1982}.
At convergence, we get the the so called ``self-consistent''  terminator (SC) \cite{Haydock:1972}
\begin{equation}
\begin{split}
   \varphi_{T}(\omega) &= [ \omega -a -b^2  \varphi_T(\omega )]^{-1} ,
   \label{t1}
\end{split}
\end{equation}
which has the solution
\begin{equation}
\begin{split}
   \varphi_{T}(\omega) &= \frac{\omega -a -  \sqrt{(\omega -a)^2- 4b^2}}{2b^2},
   \label{eq:terminator_1}                                                                                                                                                           
\end{split}
\end{equation}
where the negative root was chosen. In the TDA case we only look at positive energies, so the 
dynamical dipole polarizability could be approximated (with sufficient broadening) 
to be a continuous distribution where the terminator 
(\ref{eq:terminator_1}) can be used. For the full BSE 
case however, both positive and negative frequencies 
are explicitly treated. Since the time-ordered polarizability (as well as the case 
without any imaginary convergence factor) is symmetric 
around $\omega=0$ it has at least two distributions 
separated by a gap. 

\begin{figure}[t!]

\includegraphics[width=1.0\columnwidth, angle=0]{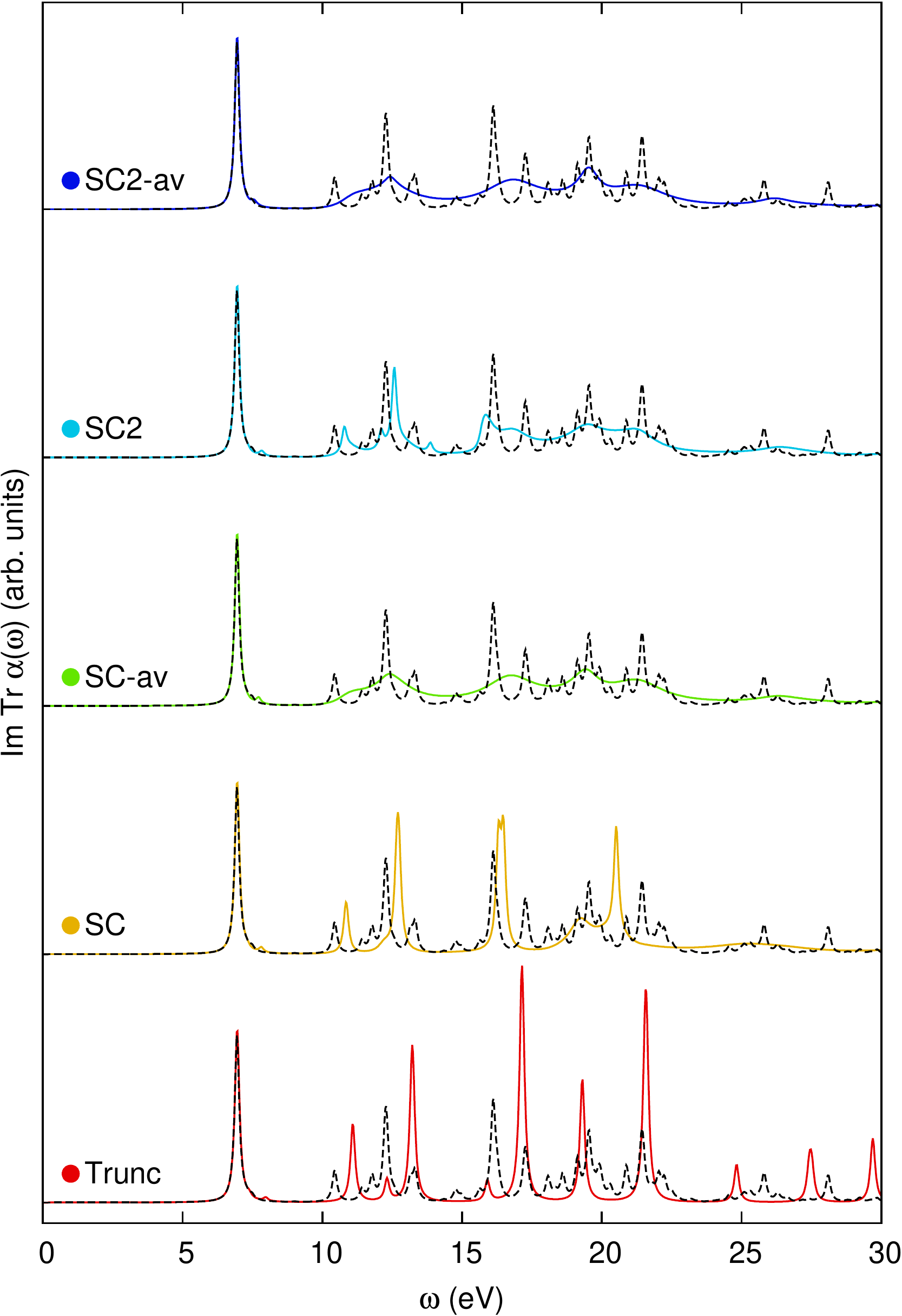}

\caption{Comparison of different terminators of the continued fraction for the iteratively computed 
TDA dynamical dipole polarizability of benzene. The number of iterations was set to
20. See the text for description 
of the different terminators.}
\label{fig:BSE_cont_frac_termination_TDA}
\end{figure}

\begin{figure}[t!]

\includegraphics[width=1.0\columnwidth, angle=0]{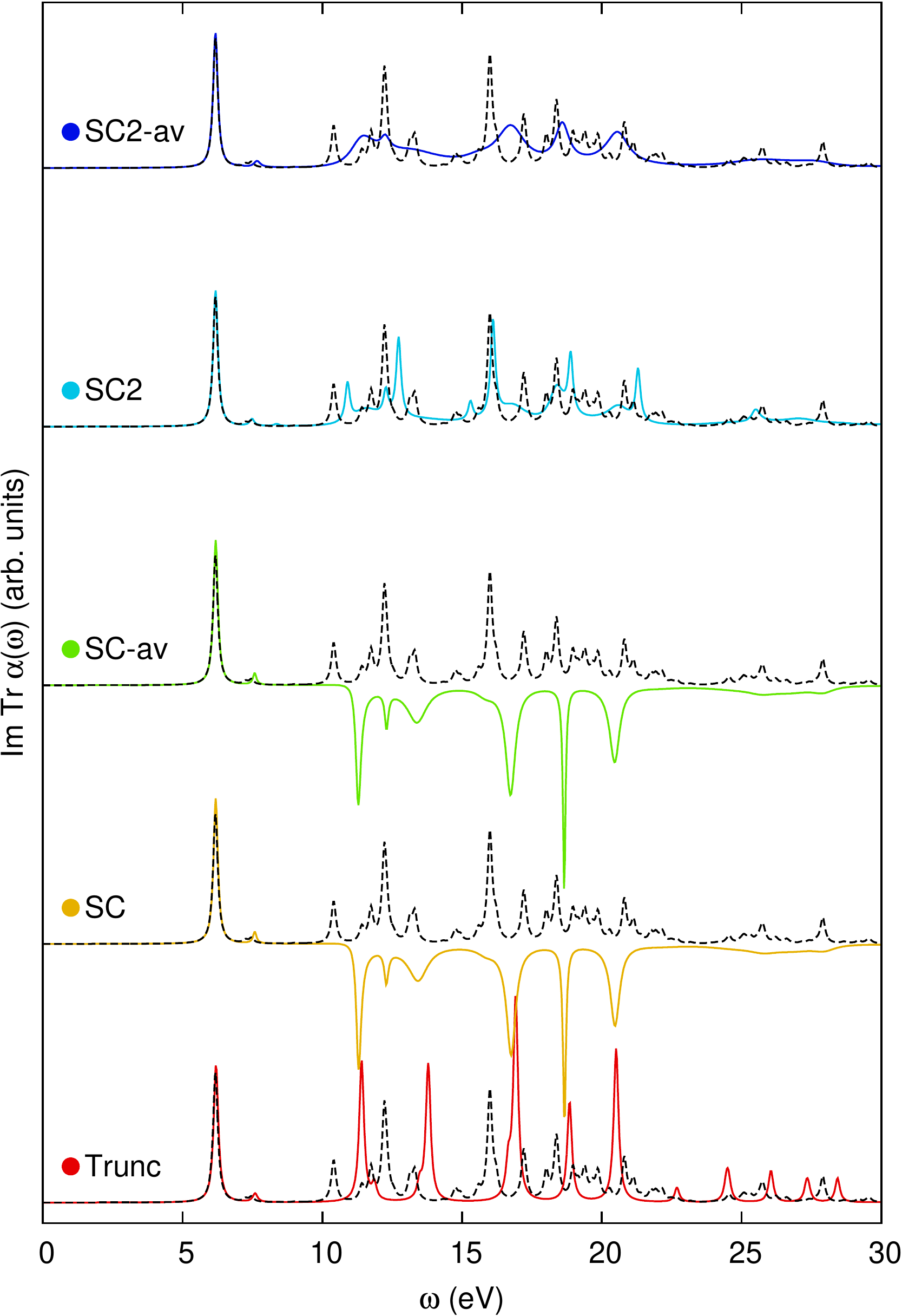}

\caption{Comparison of different terminators of the continued fraction for the iteratively computed 
full BSE dynamical dipole polarizability of benzene. The number of iterations was set to
100. See the text for description 
of the different terminators.}
\label{fig:BSE_cont_frac_termination_nonTDA}
\end{figure}

The presence of the gap in the middle of the distribution is included
 in the second model we look at.  Turchi {\it et al} analyzed the behavior 
of the recursion coefficients for densities of states with a 
gap and showed that if $2E_G$ is the gap ($a$ and $E_W$ defined 
as before) the $a_n$ coefficients oscillate with limits $a_{\pm}=a \pm E_G$, 
and $b_n$ with limits  $(b_{\pm} = E_W \pm E_G)/2$. \cite{Turchi:1982}
The period of the oscillations depends on the details of the density
of states. If no gap is present, we have the situation of equation 
(\ref{eq:terminator_1}). For a symmetric distribution around 
the middle of a single gap, which could be a good approximation 
to the full BSE case, the period is two, and the terminator is
\begin{equation}
\begin{split}
\varphi_T(\omega) = [\omega -a_{\pm} - b_{\pm}^2[\omega -a_{\mp} -b_{\mp}^2 \varphi_T(\omega)]^{-1}]^{-1}
\end{split}
\end{equation}
which has the solution (for the negative root)
\begin{equation}
\begin{split}
   \varphi_T(\omega) &= -p(\omega)/2  - \sqrt{p^2(\omega)/4 - q(\omega)}, \\
   p(\omega) &= -\frac{(\omega-a_{\pm})(\omega-a_{\mp} 
   -b_{\pm}^2 + b_{\mp}^2)}{(\omega-a_{\pm})b_{\mp}^2},\\
   q(\omega) &= \frac{\omega -a_{\mp}}{(\omega-a_{\pm}) b_{\mp}^2}.  
   \label{eq:terminator_2}   
\end{split}
\end{equation}
We denote this model SC2. Because of symmetry around frequency $\omega = 0$ 
the $a_n$ coefficients will oscillate around zero in the full BSE case. Indeed, since only the odd moments 
of the line shape contribute to $a_n$, they should be zero \cite{Haydock:1972, Mori:1965}. However, in practice,  orthogonality
between the Lanczos vectors will eventually be lost due to numerical errors, and this introduces non-zero
values of $a_n$.  
%%%%
%%%% Averaging 
%%
In practice one can at any point in the recursion sequence make 
the assumption that the coefficients have converged and so put in
the value of the last computed coefficients in equation (\ref{eq:terminator_1})
or (\ref{eq:terminator_2}).
Another option is to make the assumption that 
the coefficients will converge to the average value of the already computed
coefficients, removing some of the bias of the exact point in the chain the 
termination was made. We denote the averaged terminators by SC-av and SC2-av
when the average is applied for the terminator in (\ref{eq:terminator_1}) or 
in (\ref{eq:terminator_2}) respectively. 
When $a_n$ is set to zero in the equations 
(\ref{eq:terminator_1}), (\ref{eq:terminator_2}), the terminator reduces 
to the one used in refs \onlinecite{Rocca:2008, Gruening-Gonze:2011} (except for the signs
of $b_n^2$ and $b_{n+1}^2$) which is appropriate for the full BSE case. 
The consequence of the choice of terminator is illustrated 
in figures \ref{fig:BSE_cont_frac_termination_TDA} and  
\ref{fig:BSE_cont_frac_termination_nonTDA} for the TDA and full BSE case respectively.
The dynamical dipole polarizability was computed with 20 and 100 iterations
for TDA and full BSE correspondingly, while using simple truncation,
or terminators defined by equations (\ref{eq:terminator_1}) or
(\ref{eq:terminator_2}), with or without averaging. 
For TDA the self-consistent terminator SC
gives a slight improvement while SC2 does better, although it introduces more broadening. 
When averaging the coefficients we introduce even more broadening in the continuum part
of the dynamical dipole polarizability.

Looking at the $a_n$ and $b_n$ coefficients we see that they do not converge but  
oscillate, which is expected because our small basis set cannot give rise to 
a continuous dynamical dipole polarizability in the continuum. For an arbitrary stick-like distribution the behavior
of the coefficients is complicated. If we look at the averages of the coefficients, we see that 
$\langle a_n \rangle \approx 72$ eV which is close to the center of the spectrum, $65$ eV, as 
estimated as half the range of the $GW$ eigenvalues, while
$\langle b_n \rangle \approx 33$ eV 
which is close to a quarter of the range of the spectrum, as expected. 
Averages of the even and odd $b_n$ coefficients do not differ almost at all, 
hence the very similar appearance of the averaged versions of terminators SC and SC2. Using two following $b_n$ 
coefficients however, preserves some oscillations and gives a 
slightly better agreement to the converged spectrum.

In the non-TDA case, the SC terminator fails completely and gives 
negative intensities. Here it is clear that at
least two oscillating coefficients 
must be used for a reasonable description. $\langle a_n \rangle$ was confirmed to 
be zero, and the averages of the odd and even coefficients were seen to be 
$72$ and $64$ eV respectively. Their difference ($8$ eV) should correspond to half
the gap $E_G$, roughly $6$ eV in our calculations, estimated from the $GW$ eigenvalues. 
Here again, we observe that taking 
the average leads to a smoother spectrum that does not necessarily improve 
things from only using the last two coefficients. This is likely due to the complicated oscillations
coming from the stick-like distribution obtained with our small basis set.

\subsection{Demonstration of the low scaling with system size}

\begin{figure}[t!]

\includegraphics[width=1.0\columnwidth, angle=0]{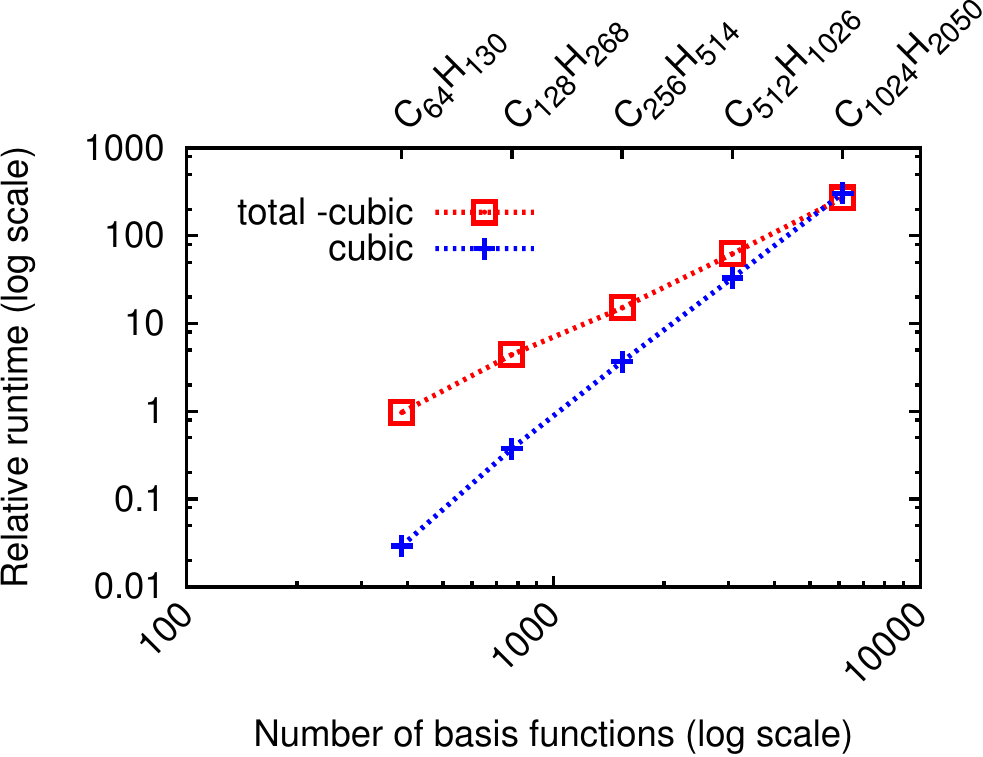}

\caption{Runtime per Lanczos vector for alkane chains of different length, divided by the runtime of the $C_{64}H_{130}$ chain. The red points  are the total runtime minus the runtime of the basis transformation, the blue the contribution of the basis transformation (denoted cubic). Dotted lines have been drawn between the points as a guide for the eye. }
\label{fig:alkane_chains}
\end{figure}

To demonstrate the scaling properties of our algorithm we performed Lanczos iterations for alkane chains of increasing length. 
One-dimensional systems are the most favorable cases for algorithms that make use of sparsity, since the number of overlapping functions will be small. To demonstrate the asymptotic scaling of our algorithm this system is also ideal --- the part that scales cubically depends on the number of AOs, while the dominant quadratic scaling operations involve the number of overlapping AOs. A sparse one-dimensional system maximizes the ratio of the former to the latter. The ground state calculation was done with SIESTA using the LDA functional and a minimal SZ basis set. Although scaling like $O(N^3)$, our $GW$ scheme turned out to 
be a bottleneck as the systems grow larger, and we therefore chose to bypass the $GW$ step and directly do a TDHF benchmark starting from LDA eigenstates. For the purpose of testing the iterative BSE algorithm the choice of starting point makes no difference. 
In figure \ref{fig:alkane_chains} we show the runtime, per Lanczos step, or alkane chains of different sizes divided by the runtime of the smallest chain, $C_{64}H_{130}$. The largest alkane chain we considered was $C_{1024}H_{2050}$ with 6146 basis functions. 
The pseudo-Hermitian algorithm was used in this comparison. 
In the figure the part of the runtime coming from the basis transformation in equations (\ref{eq:aux_vectors_mo}), (\ref{eq:aux_vectors_mo}) that should scale cubically is contrasted to the remaining runtime contributions. For small systems the basis transform is negligible in comparison to the other terms but due to its cubic asymptotic scaling it will eventually start to dominate. We see that for the largest chain considered the basis transformation consumes around half the runtime, and we would need to go to even larger systems for the cubic terms to dominate completely. 
We must here stress the fact that we have used and almost artificially sparse system in order to demonstrate the cubic scaling of the algorithm. For more realistic systems that are less sparse and have more basis functions per atom, the onset where the cubic terms start to dominate will occur much later. We can thus expect that the quadratic and lower terms will dominate for systems with up to several thousands of basis functions, i.e.,  for most systems that can be practically treated with standard DFT methods.  

\section{Conclusions}

We have described, and implemented, an iterative scheme
to compute the optical response of molecular systems 
at the Bethe-Salpeter level, using local basis sets. 
We go beyond 
the Tamm-Dancoff approximation 
by an extension of the Hermitian Haydock recursion scheme 
to the pseudo-Hermitian 
case and provide a derivation of this extension. 
We show that 
it is possible to develop an implementation with low scaling with the system size by
exploiting the localization of the basis set of numerical atomic orbitals.
Proof of principle 
calculations are shown, focusing on the case of benzene, and 
 the influence of the number of recursion vectors is discussed, as is the effect of 
different terminators of the continued fractions on the obtained dynamical dipole polarizability. 

The 
theoretical scaling of our method is $O(N^3)$. However, calculations performed for alkane 
chains containing up to 1024 carbon atoms shows that the contribution of the cubic terms is small. 
Even for the largest systems considered the contribution of the cubic terms is of comparable magnitude to
to that of the quadratic terms coming from the application of the Coulomb kernel in the atomic basis.

What we have presented here is a proof of principles of the method, plus an analysis of the convergence
of our iterative BSE scheme. Our final goal, however,  is to create a method (implemented in a suite of programs) 
capable of accurately investigating complex systems containing thousands of atoms. In order to reach this goal 
we are currently investigating ways to improve the performance of the method.
These include an efficient parallelization 
scheme and a more optimized basis set for the expansion of atomic orbital products.  
It is also the case that the $GW$ calculation needed to
obtain the quasiparticle energies 
and states, as well as the screened interaction matrix elements, 
can benefit from similar improvements.

\begin{acknowledgments}

We acknowledge support from the Deutsche Forschungsgemeinschaft (DFG) through the SFB 1083 project, 
the ANR ORGAVOLT project and the 
Spanish MINECO MAT2013-46593-C6-2-P project. 
Discussions with Mark Casida are gratefully 
acknowledged. 
DF thanks Lorin X.Benedict for discussions concerning the scaling 
of the algorithm. 
PK acknowledge financial support from the Fellows
Gipuzkoa program of the Gipuzkoako Foru Aldundia through
the FEDER funding scheme of the European Union, ÒUna
manera de hacer EuropaÓ.
FF acknowledges support from the EXTRA programme of the "Universit\`{a} degli Studi di Milano-Bicocca"
and from the Erasmus Placement programme for student mobility.
\end{acknowledgments}

\appendix

\section{Derivation of the Bethe-Salpeter equation}
\label{sec:appendix_derivation_BSE}

Here we derive the BSE equation following an approach similar to that given 
in refs \onlinecite{Bruneval_thesis:2005, Sottile_thesis:2003}. 
The purpose of this appendix is to derive the equations using our notation to 
avoid possible confusions with different notations and conventions that can be 
found in the literature.

The reducible two-point polarizability is the response of the density 
to a local perturbation $U$
\begin{equation}
\chi(\mathbf 1,\mathbf 2) = \frac{\delta \rho(\mathbf 1)}{\delta U(\mathbf 2)} = -\text{i}
\frac{\delta G(\mathbf1,\mathbf1^+)}{\delta U(\mathbf 2)}.
\label{chi-def}
\end{equation}
where in the "$+$" superscript denotes the addition of a positive infinitesimal to the time argument.
A generalization can be made to the nonlocal response of the interacting Green's function 
$G$ to a nonlocal perturbation, giving the four-point polarizability
\begin{equation}
L(\mathbf 1,\mathbf 2,\mathbf 3,\mathbf 4) =  -\text{i} \frac{\delta G(\mathbf 1,\mathbf 2)}{\delta U(\mathbf 3,\mathbf 4)} \, .
\label{pol-def}
\end{equation}
Comparing equation (\ref{chi-def}) with (\ref{pol-def}), we 
conclude that $\chi(\mathbf 1,\mathbf 2) = L(\mathbf 1,\mathbf 1^+,\mathbf 2,\mathbf 2)$.  Using the 
Schwinger functional derivative method \cite{Martin_Schwinger:1959, Hedin:1965} and references therein) the following relation can be proved
\begin{equation}
\begin{split}
L(\mathbf 1,\mathbf 2,\mathbf 3,\mathbf 4) = iG(\mathbf 1,\mathbf 4,\mathbf 2,\mathbf 3)  -iG(\mathbf 1,\mathbf 2)G(\mathbf 4,\mathbf 3) \, , 
\end{split}
\end{equation}
where the two-particle Green's function is defined as
\begin{equation}
\begin{split}
G(\mathbf 1,\mathbf 2,\mathbf 3, \mathbf 4) = (-i)^2 \langle N | \mathcal{T} \{ \hat \psi(\mathbf 1)\hat \psi(\mathbf 2) \hat \psi^{\dagger}(\mathbf 4)\hat \psi^{\dagger}(\mathbf 3)   \} | N \rangle \, .
\end{split}
\end{equation}

Instead of working with the two-particle Green's function we will directly 
derive the Bethe-Salpeter equation for the four-point polarizability $L$. We 
will use two relations: the chain rule
\begin{equation}
\frac{\delta F[G[H]](\mathbf 1,\mathbf 2) }{\delta H(\mathbf 3,\mathbf 4)} = 
\int d(\mathbf 5\mathbf 6) \frac{\delta F[G](\mathbf 1,\mathbf 2) }{\delta G(\mathbf 5,\mathbf 6)}  
\frac{\delta G[H](\mathbf 5,\mathbf 6) }{\delta H(\mathbf 3,\mathbf 4)} \, ,
\label{chain-rule}
\end{equation}
and a transformation of a derivative of a function to include its inverse
\begin{equation}
\frac{\delta F(\mathbf 1,\mathbf 2) }{\delta G(\mathbf 3,\mathbf 4)} = -
\int d(\mathbf 5\mathbf 6) F(\mathbf 1,\mathbf 5)  \frac{\delta F^{-1}(\mathbf 5,\mathbf 6) }{\delta G(\mathbf 3,\mathbf 4)} F(\mathbf 6,\mathbf 2).
\label{eq:derivative_inverse}
\end{equation}
Using equation (\ref{eq:derivative_inverse}) we can write
\begin{equation}
\frac{\delta G(\mathbf 1,\mathbf 2)}{\delta U(\mathbf 3,\mathbf 4)}  = -\int d(\mathbf 5\mathbf 6) G(\mathbf 1,\mathbf 5)G(\mathbf 6,\mathbf 2)  
\frac{\delta G^{-1}(\mathbf 5,\mathbf 6) }{\delta U(\mathbf 3,\mathbf 4)} \, .
\label{eq:dG_dU}
\end{equation}
From the Dyson equation for interacting Green's function $G$
we have
\begin{equation}
G^{-1} (\mathbf 5,\mathbf 6)=  G_0^{-1}(\mathbf 5,\mathbf 6) - U(\mathbf 5,\mathbf 6) - v_\text{H}(\mathbf 5)\delta(\mathbf 5,\mathbf 6) -\Sigma(\mathbf 5,\mathbf 6),
\end{equation}
where we added the external potential $U$ to the Hamiltonian 
(it will be set to zero after the derivatives have been taken) 
and the Hartree potential $v_\text{H}$ is taken outside of 
the non-interacting Green's function $G_0$. 
Evaluating the functional derivative, remembering that 
$G_0$ is independent of $U$, we get
\begin{equation}
\begin{split}
\frac{\delta G^{-1}(\mathbf 5,\mathbf 6) }{\delta U(\mathbf 3,\mathbf 4)} &= 
-\delta(\mathbf 3,\mathbf 5)\delta(\mathbf 4,\mathbf 6) - \frac{\delta}{\delta U(\mathbf 3,\mathbf 4)} 
\left [ v_\text{H}(\mathbf 5)\delta(\mathbf 5,\mathbf 6) +\Sigma(\mathbf 5,\mathbf 6) \right ] \\
&= -\delta(\mathbf 3,\mathbf 5)\delta(\mathbf 4,\mathbf 6) - \int d(\mathbf 7\mathbf 8) \frac{\delta}{\delta G(\mathbf 7,\mathbf 8)} 
 [ v_\text{H}(\mathbf 5)\delta(\mathbf 5,\mathbf 6) \\
&+\Sigma(\mathbf 5,\mathbf 6)  ]  \frac{\delta G(\mathbf 7,\mathbf 8)}{\delta U(\mathbf 3,\mathbf 4)},
\label{eq:dGinv_dU}
\end{split}
\end{equation}
where in the last step we used the chain rule (\ref{chain-rule}).
Combining equations 
(\ref{eq:dG_dU}, \ref{eq:dGinv_dU}) we obtain
\begin{equation}
\begin{split}
\frac{\delta G(\mathbf 1,\mathbf 2)}{\delta U(\mathbf 3,\mathbf 4)}  =& G(\mathbf 1,\mathbf 3)G(\mathbf 4,\mathbf 2)  \\
&+ \int d(\mathbf 5\mathbf 6\mathbf 7\mathbf 8) G(\mathbf 1,\mathbf 5)G(\mathbf 6,\mathbf 2)  \frac{\delta}{\delta G(\mathbf 7,\mathbf 8)} 
[ v_\text{H}(\mathbf 5)\delta(\mathbf 5,\mathbf 6) \\
& +\Sigma(\mathbf 5,\mathbf 6)  ]  \frac{\delta G(\mathbf 7,\mathbf 8)}{\delta U(\mathbf 3,\mathbf 4)} \, .
\end{split}
\end{equation}
Defining
\begin{align}
L_0(\mathbf 1,\mathbf 2,\mathbf 3,\mathbf 4) &= -\text{i} G(\mathbf 1,\mathbf 3)G(\mathbf 4,\mathbf 2) \, ,  \\ 
K(\mathbf 5,\mathbf 6,\mathbf 7,\mathbf 8) &=  \text{i} \frac{\delta}{\delta G(\mathbf 7,\mathbf 8)} \left [ 
v_\text{H}(\mathbf 5)\delta(\mathbf 5,\mathbf 6) +\Sigma(\mathbf 5,\mathbf 6) \right ]  \, ,
\end{align}
we finally get the Bethe-Salpeter equation
\begin{equation}
\begin{split}
L(\mathbf 1,\mathbf 2,\mathbf 3,\mathbf 4)  =& L_0(\mathbf 1,\mathbf 2,\mathbf 3,\mathbf 4) \\
+ &\int d(\mathbf 5\mathbf 6\mathbf 7\mathbf 8)L_0(\mathbf 1,\mathbf 2,\mathbf 5,\mathbf 6) K(\mathbf 5,\mathbf 6,\mathbf 7,\mathbf 8)L(\mathbf 7,\mathbf 8,\mathbf 3,\mathbf 4)  \, .
\end{split}
\label{eq:BSE_time}
\end{equation}
Up to this point the derivation has been exact. In order to obtain the working expression for the  BSE kernel, $K$, we now make use of 
the $GW$ approximation to the self energy. In this case both the Hartree potential $v_\text{H}$ and the self energy $\Sigma$ 
can be expressed in terms of $G$:

\begin{align}
v_\text{H}(\mathbf 1) &= \int d(\mathbf  2) v(\mathbf 1,\mathbf 2)\rho(\mathbf 2)=  -\text{i} \int d(\mathbf 2) v(\mathbf 1,\mathbf 2) G(\mathbf 2,\mathbf 2^+) \, ,\\
\Sigma(\mathbf 1,\mathbf 2) &= \text{i} G(\mathbf 1,\mathbf 2) W(\mathbf 1,\mathbf 2) \, ,
\end{align}
which, neglecting the dependence of $W$ on $G$, gives
\begin{equation}
K(\mathbf 1,\mathbf 2,\mathbf 3,\mathbf 4)  = v(\mathbf 1,\mathbf 3)\delta(\mathbf 1,\mathbf 2)\delta(\mathbf 3,\mathbf 4) - W(\mathbf 1,\mathbf 2)\delta(\mathbf 1,\mathbf 3)\delta(\mathbf 2,\mathbf 4)\, .
\end{equation}

Here we note that the bare Coulomb interaction is instantaneous $v(\mathbf 1, \mathbf 2) = v(1,2)\delta(t_2-t_1)$, but this is not in general the case for $W$. Equation (\ref{eq:BSE_time}) still depends on four times. 
For our purposes, we want to look at the 
response at time $t$ from a perturbation at time $t'$, that is our 
perturbations are local in time. In this case we can express $L$ in terms of the 
density matrix $\rho(1,2,t)$ as
\begin{equation}
\begin{split}
 L(\mathbf 1,\mathbf 2,\mathbf 3,\mathbf 4) = \frac{\delta \rho(1,2,t_1)}{\delta U(3,4,t_3)} \delta(t_1-t_2) \delta(t_3-t_4) \, ,
  \end{split}
  \label{eq:L0_rho}
\end{equation}
where we identify $t=t_1$ and $t'=t_3$. Since the initial time is arbitrary 
for a system in equilibrium 
--- the state of the system does not change
in time when we are in the ground state ---
we furthermore only have to worry about the difference
$t'-t$.  As in the Dyson equation for the Green's function $G$, we can
then Fourier transform to get a dependence 
of only one frequency, thus giving
\begin{equation}
\begin{split}
&L(1,2,3,4 \, | \, \omega)  = L_0(1,2,3,4 \, | \,  \omega) \\
&+ \int d(5678)L_0(1,2,5,6 \, | \,  \omega) K(5,6,7,8  \, | \,  \omega)L(7,8,3,4 \, | \,  \omega)  \, .
\end{split}
\label{eq:BSE_final}
\end{equation}
\section{Spin structure of the effective BSE and dependence of the occupations for the polarizability}
\label{sec:appendix_spin}

\subsubsection{Spin structure}

The BSE Hamiltonian can be written in matrix form as
\begin{equation}
\begin{split}
H^\text{BSE} =H^0 + F K % (H^\text{ex} + H^\text{dir} )
\end{split}
\end{equation}
with $K= H^\text{ex} + H^\text{dir}$.
We assume a singlet closed shell ground state so 
the spatial orbitals are the same for spin up and 
spin down. Explicitly writing out the spin 
dependence of the orbitals as $\psi_i(1) = 
\psi_i(\bm{r})x_i(\sigma) $, and $\psi_i(2) = 
\psi_i(\bm{r'})x_i(\sigma') $, where the spin wave  function $x_i(\sigma)$ 
can be either $\alpha(\sigma)$ or $\beta(\sigma)$. 
Due to orthogonality of the spin wave functions we get
\begin{equation}
\begin{split}
H^0_{ij,kl} &= (\epsilon_j-\epsilon_i) \delta_{ik} \delta_{jl} \delta_{x_i x_k}
 \delta_{x_j x_l} \, ,\ \\
H^\text{ex}_{ij, kl} &=\int d^3r d^3r' \psi_i^*(\bm{r}) 
\psi_j(\bm{r}) v(\bm{r},\bm{r'}) \psi_k(\bm{r'}) \psi_l^*(\bm{r'}) 
 \delta_{x_i x_j} \delta_{x_k x_l} \, ,\\
H^\text{dir}_{ij, kl} &= -\int d^3r d^3r' \psi_i^*(\bm{r}) 
\psi_k(\bm{r}) W(\bm{r},\bm{r'}) \psi_j(\bm{r'}) 
\psi_l^*(\bm{r'})  \delta_{x_i x_k} \delta_{x_j x_l} \, ,
\end{split}
\end{equation}
This gives the following structure of the problem in the spin indices
\begin{widetext}
\begin{equation}
\bordermatrix{%
 & \alpha \alpha & \beta \beta & \alpha \beta & \beta \alpha \cr
 \alpha \alpha & H^0  + F(H^\text{ex}  + H^\text{dir}) & FH^\text{ex}  & 0 & 0  \cr
  \beta \beta  &  FH^\text{ex}  & H^0  + F(H^\text{ex}  + H^\text{dir})  & 0 & 0 \cr
 \alpha \beta & 0 & 0 & H^0  + FH^\text{dir}  & 0  \cr
 \beta \alpha & 0 & 0 & 0 & H^0  + FH^\text{dir}  \cr }.
\end{equation}
\end{widetext}
The upper left 2 x 2 block can easily be diagonalized  to give
\begin{widetext}
\begin{equation}
\bordermatrix{%
 & \frac{1}{\sqrt{2}}(\alpha \alpha + \beta \beta) &  \frac{1}{\sqrt{2}}(\alpha \alpha - \beta \beta)  & \alpha \beta & \beta \alpha \cr
\frac{1}{\sqrt{2}}(\alpha \alpha + \beta \beta) & H^0 + F(2H^\text{ex} + H^\text{dir}) & 0 & 0 & 0  \cr
 \frac{1}{\sqrt{2}}(\alpha \alpha - \beta \beta)  &  0 & H^0 + FH^\text{dir} & 0 & 0 \cr
 \alpha \beta & 0 & 0 & H^0 + FH^\text{dir} & 0  \cr
 \beta \alpha & 0 & 0 & 0 & H^0 + FH^\text{dir} \cr } \, ,
\end{equation}
\end{widetext}
which leads to one singlet solution, where $H^\text{ex}$ is included with a factor of 2,
and three triplet solutions where $H^\text{ex}$ is absent. Knowing this, we work just with 
the real space quantities, remembering to include the correct scaling factor in front 
of $H^\text{ex}$ depending on if we want a singlet or a triplet solution 
\begin{equation}
\begin{split}
K^\text{singlet}_{ij,kl} &= 2H^\text{ex}_{ij,kl} +H^\text{dir}_{ij,kl} \, ,\\
K^\text{triplet}_{ij,kl} &= H^\text{dir}_{ij,kl} \, .\\
\end{split}
\end{equation}
For the dipole elements we have
\begin{equation}
\begin{split}
D_{ij}^{m, \text{singlet}} &= \sqrt{2} \int d^3r 
\psi_i(\bm{r} )^* \bm{r}_{m} \psi_j(\bm{r}) \, , \\
D_{ij}^{m, \text{triplet}} &= 0 \, .
\end{split}
\end{equation}

\subsubsection{Occupation number structure}
The time-ordered four-point polarizability 
\begin{equation}
\begin{split}
L_{ij,kl}(\omega) = [(\omega  +\text{i}\gamma(f_{i'}-f_{j'}))\delta_{i'k'}\delta_{j'l'} 
- H^\text{BSE}_{i'j',k'l'} ]^{-1}_{ij,kl} (f_k-f_l)  \, ,
\end{split}
\end{equation}
can be written in matrix form as
\begin{equation}
\begin{split}
L(\omega) = [(\omega  +\text{i}\gamma F)I   - H^\text{BSE} ]^{-1} F  \, .
\end{split}
\label{eq:L_inverse_H_t}
\end{equation}
From this expression it looks like we have to use all pairs, that is not 
only particle-hole and hole-particle pairs but also particle-particle and 
hole-hole pairs. But actually, only the  particle-hole and hole-particle  
pairs contribute to $L$.  To see this we set up the $H^\text{BSE}$, and $F$ matrices 
in blocks corresponding to the $\{vc\}$,  $\{cv\}$,$\{vv\}$ and $\{cc\}$ sectors
\begin{equation}
\begin{split}
&H^\text{BSE} =
 \bordermatrix{%
 & vc & cv & vv & cc \cr
 vc & H^{0}+ K & K &  K & K \cr
 cv & -K& H^{0} - K &  -K & -K\cr
 vv &  0 & 0 & H^{0} & 0 \cr  
 cc &  0 & 0 & 0 & H^{0} \cr 
 }   \, ,
\end{split}
\label{}
\end{equation}
%
%and $F$
%
\begin{equation}
\begin{split}
&F =
 \bordermatrix{%
& vc & cv & vv & cc \cr
 vc & I & 0 &  0 & 0 \cr
 cv & 0 & -I &  0 & 0 \cr
 vv &  0 & 0 & 0 & 0 \cr  
 cc &  0 & 0 & 0 & 0 \cr 
 }    \,.
\end{split}
\label{}
\end{equation}
This gives the following matrix to be inverted in equation (\ref{eq:L_inverse_H_t})
\begin{widetext}
\begin{equation}
\begin{split}
(\omega  + \text{i} \gamma F)I- H^\text{BSE} =
\bordermatrix{%
& vc & cv & vv & cc \cr
 vc & (\omega + \text{i} \gamma) I -(H^{0}+ K) & -K &  -K & -K \cr
 cv & K & ( \omega  - \text{i} \gamma) I  -(H^{0} -K) &  K & K \cr
 vv &  0 & 0 &  \omega I -H^{0} & 0 \cr  
 cc &  0 & 0 & 0 &  \omega I -H^{0} \cr 
 }   \, . 
\end{split}
\end{equation}
\end{widetext}
The inverse of a matrix with this block structure is
\begin{equation}
\begin{split}
\left( \begin{array}{cc}
 A & B\\
 0 & D 
\end{array} \right)^{-1} =%
\left( \begin{array}{cc}
 A^{-1} & -A^{-1}BD^{-1}\\
 0 & D^{-1} 
\end{array} \right) \, .
\end{split}
%\label{eq:BSE}
\end{equation}
Looking at the polarizability $L = [(\omega  + \text{i} \gamma F)I-H^\text{BSE}]^{ -1}F $ 
we see that due to the 
leftmost $F$ matrix only the $A$-block, that is the $\{vc\}$ and $\{cv\}$-sectors of $H^\text{BSE}$, contribute to 
$L$. Diagonalizing $H^\text{BSE}$ and expanding in left and right eigenvectors
 gives the following expression
\begin{equation}
\begin{split}
L_{ij,kl}(\omega) &=  \sum_{\lambda, \lambda'}
\frac{A^{\lambda}_{ij}  S^{-1}_{\lambda, \lambda'} A^{\lambda' *}_{kl} 
(f_k-f_l) }{\omega  -\epsilon_{\lambda} +\text{i}\gamma(f_i-f_j)  }  \, .
\end{split}
\label{eq:L_t}
\end{equation}
Note that this is the time-ordered polarizability, the retarded one that we need 
for the response, that is equation (\ref{eq:L_r}), is obtained by setting  the sign of the small imaginary part in 
the denominator to always be positive. We can also use the relations  
$\operatorname{Im} L^t(\omega) = \text{sgn}(\omega) \operatorname{Im} L^r (\omega) $ and  
$\operatorname{Re} L^t(\omega) = \operatorname{Re} L^r (\omega) $, 
where the superscripts "$t$" denotes time-ordered and "$r$" retarded.

%\bibliography{Refs}

\end{document}